# Power System Capacity Planning Considering Seasonal Hydrogen Storage by Salt Caverns

Xueqian He[1], Tianguang Lu[1,2*], Jing Li[1], Wanxing Sheng[3], Rui Li[3]

[1]School of Electrical Engineering, Shandong University, Jinan 250061, China
[2]School of Engineering and Applied Sciences and Harvard China Project, Harvard University, Cambridge, MA 02138, United States
[3] National Power Distribution Center, State Grid Corporation of China, Beijing 100031, China
* tlu@sdu.edu.cn

**Abstract:** In China, air conditioning in summer and electric heating in winter lead to seasonal volatility in load power. Therefore, it is urgent to develop economic and efficient long-term energy storage systems to enhance peak regulation. Power-to-hydrogen technology is a perspective solution to balance seasonal power fluctuation. However, current hydrogen storage methods have shortcomings such as small storage capacity, high levelized cost and low operation safety, which the salt cavern hydrogen storage could overcome. This paper considers the use of hydrogen storage in salt caverns as a means of peak shaving. To minimize the overall operating cost, a comprehensive power system capacity planning model is proposed with the consideration of hydrogen storage in salt caverns, which is implemented by adopting an improved fast unit commitment method. Considering the seasonal characteristics of the load power in Jiangsu Province, the capacity of the power system in 2050 has been planned. According to the case study, after the optimal deployment of the salt cavern hydrogen storage system (SCHSS), the construction capacity of renewable units (especially wind power) will be significantly increased with environmental friendliness and lower costs. Compared with the current energy storage method, the overall cost of SCHSS-incorporated power system will be reduced by 22.2% with a carbon emission reduction of 24.4%, and the amount of curtailed wind and solar power will be reduced by 27.0% and 13.6%, respectively.

**Nomenclature**

| | | | |
|---|---|---|---|
| $x_j(t)$ | The on/off status of unit j | $a_{\text{salt}}$ | The amortized investment cost of salt caverns |
| $s_j(t)$ | The startup behavior of the unit | $I_{\text{salt}}$ | The newly-built capacity of salt caverns |
| $u_j(t)$ | The shutdown behavior for the unit | $r_{\text{P2H}}$ | The operation and maintenance cost of electrolytic hydrogen production units |
| $p_j(t)$ | The power output for unit j at time t | $r_{\text{H2P}}$ | The operation and maintenance cost of fuel cell units |
| $p_j^{max}$ | The nameplate capacity of unit j | $I_{\text{P2H}}$ | The capacity of electrolytic hydrogen production units |
| $\overline{\delta_j}$ , $\underline{\delta_j}$ | The ratios of the maximum and minimum power output of unit j to its nameplate capacity | $I_{\text{H2P},t}$ | The capacities of fuel cell units |
| $\widehat{P_t^i}$ | The online capacity of group $i$ at time t (discrete variable) | $\alpha_t$ , $\beta_t$ | The hourly capacity factors of wind and solar power at time t |
| $\widehat{S_t^i}$ , $\widehat{U_t^i}$ | The startup and shutdown capacity of group $i$ at time t (discrete variable) | $\overline{\mu}_i$ , $\underline{\mu}_i$ | The maximum and minimum output ratios of thermal units of type $i$ |
| $\overline{p}_t^i$ | The online capacity of unit group $i$ at time t (continuous variables) | $RU^i$, $RD^i$ | The upward and downward ramping ratios of thermal units of type $i$ |
| $s_t^i$, $u_t^i$ | The startup and shutdown capacity of group $i$ at time t (continuous variables) | $VS^i$, $VD^i$ | The start-up and shut-down ramp limits of thermal units of type $i$ |
| $p_t^i$ | The power output of thermal units of type $i$at time t | $UT^i$, $DT^i$ | The minimum start-up and shut-down time of thermal units of type $i$ |
| $p_t^w$ | The power output for wind at time t | $p_t^e$ | The generating power of CHP units at time t |
| $p_t^s$ | The power output for solar at time t | $p_t^h$ | The heat generated by CHP units at time t |
| $S_i$ | The total capacity of unit group $i$ | $TE$ | The thermoelectric ratio of CHP units |
| $W_a$ | The investment cost of new unit | $HM_t$ | The heat demand at time t |
| $W_f$ | The operation and maintain cost of thermal units and variable renewable energy units | $DM_t$ | The load demand at time t |
| $W_v$ | Both the start-up cost and the fuel cost for thermal units | $p_{\mathrm{H2P},t}$ | The power consumed by electrolytic hydrogen production at time t |
| $W_h$ | The cost of SCHSS | $p_{\mathrm{P2H},t}$ | The power consumed by hydrogen power generation at time t |
| $a^i$ | The amortized investment cost for thermal units of type $i$ | $p_{\mathrm{P2H},max}$ | The maximum power of hydrogen production by electrolysis at time t |
| $a^w$ | The amortized investment cost for wind units | $H_{\mathrm{H2P},max}$ | The maximum hydrogen consumption of fuel cells at time t |
| $a^s$ | The amortized investment cost for solar units | $H_{\mathrm{P2H},t}$ | The hydrogen production at time t |
| $f^i$ | The operation and maintain cost of thermal units of type $i$ | $H_{\mathrm{H2P},t}$ | The hydrogen consumption at time t |
| $f^w$ | The operation and maintain cost for wind units | $\eta_{\mathrm{P2H}}$ | The hydrogen production efficiency of electrolysis |
| $f^s$ | The operation and maintain costs for solar units | $\eta_{\mathrm{H2P}}$ | The power generation efficiency of fuel cell units |
| $I_{\mathrm{new}}^i$ | The newly-built capacity of thermal units of type $i$ | $H_{H,\mathrm{in},t}$ | The amount of hydrogen filled into salt caverns at time t |
| $I_{\mathrm{new}}^w$ | The newly-built capacity for wind units | $H_{H,\mathrm{out},t}$ | The amount of hydrogen extracted from salt caverns at time t |
| $I_{\mathrm{new}}^s$ | The newly-built capacity for solar units | $L_H$ | The ratio of hydrogen leakage to hydrogen storage in salt caverns per hour |
| $I^i$ | The total capacity of thermal units of type $i$ | $S_{H,t}$ | The hydrogen storage capacity of salt caverns at time t |
| $I^w$ | The total capacity for wind units | $S_{H,\mathrm{ini}}$ | The initial hydrogen storage capacity of salt caverns |
| $I^s$ | The total capacity for solar units | $S_{H,max}$ | The maximum hydrogen storage capacity of salt caverns |
| $c^i$ | The fuel cost for thermal units of type $i$ | $\eta_{H,\mathrm{in}}$ , | The efficiency of charging hydrogen into salt caverns |
| $SD^i$ | The start-up cost for thermal units of type $i$ | $\eta_{H,\mathrm{out}}$ | The efficiency of extracting hydrogen from salt caverns |

## 1. INTRODUCTION

In recent years, with the rapid development of renewable energy technology, the proportion of renewable energy in power systems has increased rapidly[1, 2]. By the end of December 2022, the installed capacity of wind power in China is about 370 million kilowatts, and that of solar power is about 390 million kilowatts, both ranking first in the world. The total installed capacity of renewable generation in China has accounted for more than 34% of the world [3-6]. However, the load of air conditioning in summer and electric heating in winter has led to the obvious seasonal volatility of the annual electric power of China, which makes peak regulation of the power system difficult [7]. According to the forecast, there will be about 90 million kW peak-shaving capacity gap in the operation area of the State Grid of China in 2030, which poses a significant challenge to the safe and stable operation of the power systems [8]. In addition, seasonal loads will cause power load demand to not match the output of variable renewable energy. This can also make it difficult for renewable energy to be consumed. According to the statistics of the National Energy Administration of China, 20.61 billion kilowatt-hours of wind power and 6.78 billion kilowatt-hours of solar power were curtailed in 2021[9-11]. The curtailment rate of wind and solar in China far exceeds other developed countries around the world. In 2015, Germany had already controlled the curtailment rate of wind and solar to less than 1% [12, 13]. However, the average wind and solar curtailment rates in China were still 3.1% and 1.8% respectively in 2021[14]. In particular, the curtailed wind rate in Qinghai Province reached 13.8%, and the curtailed solar rate in Tibet exceeded 19% [15, 16].

Hydrogen energy is regarded as the most promising clean energy in the 21st century recognized by the international community, as the global development of hydrogen energy is entering a stage of rapid industrialization [17]. The annual hydrogen production of China ranks first in the world reaching about 33 million tons. As an essential bridge between the preparation and utilization of hydrogen, hydrogen storage technology plays a strong part. However, as a traditional technology, the above-ground hydrogen storage method such as hydrogen storage tanks, sustain high energy storage costs. Taking the cost of land purchase and related infrastructure construction out of the equation, the unit capacity cost of hydrogen storage tanks has reached $15/kWh [18].

Moreover, the capacity and the injection and extraction volume per unit time flow of the above-ground hydrogen storage system is limited. For instance, the average capacity of each hydrogen storage tank of each hydrogen refueling station of the Shanghai World Expo is only 24 kg, and the flow rate of injecting and extracting hydrogen should not exceed 5 kg/min [19]. It is difficult for hydrogen storage tanks to satisfy large-scale and long-term energy storage needs [20]. In addition, the safety of the above-ground hydrogen storage system is poor due to a high frequency of hydrogen leakage, combustion, explosion and other accidents [21-23]. Since 2019, more than five large-scale hydrogen storage tank explosions have appeared that caused severe consequences and caught global attention. Besides, small -scale hydrogen storage tank explosion accidents are innumerable.

Compared with the above-ground hydrogen storage method, the underground hydrogen storage (UHS) has the advantages of low equipment investment, low construction cost, large reserves, and broad adaption to both short-term and long-term operation scenarios [24, 25]. Moreover, it has good performance of tightness, low leakage rate and low pollution risk. In addition, the hydrogen storage devices are built underground, which can save land resources [26-28]. UHS is an ideal hydrogen storage method, for which many media can be used, including artificial underground spaces, porous rocks, etc. [29] However, if porous rocks such as depleted oil and gas deposits are used for hydrogen storage, hydrogen may react with the mineral components in the reservoir. This will not only consume part of the stored hydrogen, but also produce reactions to block the pores of the reservoir, which is not conducive to long -term storage of hydrogen [30]. Salt rock is recognized as an ideal hydrogen energy storage medium due to its extremely low permeability and good creep performance [31, 32]. The special chemical properties of salt ensure the stability and sealing of energy storage. Salt caverns are artificial cave formed in the hollowed salt mine, which are suitable for storing high -pressure gases [33-35]. At present, the world has launched an attempt to use salt caverns for hydrogen storage, but they are still on a very early stage. Twelve units in seven countries, including Germany, France and the United Kingdom, have jointly launched an underground hydrogen storage project, "HyUnder", and have evaluated the long-term and large-scale hydrogen storage potential of the underground salt caverns in Germany, Britain, France, the Netherlands, Romania and Spain[36, 37]. A total of four salt cavern hydrogen storage projects have been built in the world, which are Clemens Dome, Moss Bluff, Spindletop and Tessside [38]. However, the research on the underground hydrogen storage in China lags, with the large-scale SCHSS deployments remaining blank [39-41].

Current research on hydrogen storage in salt caverns by scholars can be divided into two directions. One is the assessment of the hydrogen storage potential of salt caverns in a specific area. Dilara Gulcin Caglayan et al. evaluated the underground salt structure based on the scale of European salt mine and the storage capacity of salt caverns in [42]. They estimated the storage capacity of individual caves based on thermodynamic considerations from site-specific data. In reference [43], scholars proposed a method to evaluate the hydrogen storage potential of salt caverns by comprehensively considering the size and depth of the selected salt mine. This method can calculate the possibility to build the salt cavern of a specific depth and volume in the selected salt mine. Reference [44] studies the capacity estimation of the hydrogen storage based on geological resources. A consistent approach is presented to identify specific areas that may be suitable for the development of hydrogen storage salt caverns for a given use. The other research direction is the study of power system planning considering salt cavern energy storage. In [45], Lukas

Weimann studied the optimal hydrogen production in the zero-carbon emission renewable energy system including underground hydrogen storage. An optimization method of completely renewable and zero - emission energy system is presented to satisfy the power and hydrogen needs of the Netherlands. In order to study the capacity allocation and operation mode of the future power-hydrogen energy system, [46] established a hydrogen energy system collaborative optimization planning model considering hydrogen storage in salt caverns. The model was applied in Zhejiang Province, China, to optimize the zero-carbon path for the coordinated development of the power-hydrogen energy system. Sheila Samsatli simulated the optimal design and operation of the wind-hydrogen-electricity integrated network in [47]. The optimal configuration of wind-hydrogen-electric system including salt cavern hydrogen storage system is studied. The model determines the optimal number, size and location of each equipment, the hourly operation of each technology and so on. [48] proposed a two-stage planning method for the hydrogen penetrated energy system. And a detailed hydrogen penetrated energy system model is established. By conducting case studies for typical regions of China, it is demonstrated that the hydrogen penetrated energy system can lower the cost and reduce carbon dioxide emissions.

According to the literatures, most researches on hydrogen storage in salt caverns only focuses on the capacity planning of the energy storage systems, which ignore the long-term operation effects and comprehensive capacity planning for power systems. In addition, most of the current researches on power systems including SCHSS lack quantitative economic analysis.

To medicate the above shortcomings, this paper considers the electricity-to-hydrogen conversion technology and the SCHSS to establish a power system capacity planning model with the goal of minimizing the operating cost of the power system. In this paper, the long-term optimization of the power capacity is carried out, and the capacity of different units is optimized with hour as the time granularity. This article also concludes with an economic analysis of the planning results. Jiangsu Province has significant seasonal load characteristics, and its power consumption and salt cavern scale are among the top in China, making it the best province to test the effect of SCHSS. In this paper, Jiangsu Province is taken as an example for case analysis, where the seasonal characteristics of the load of Jiangsu Province is analyzed and the regulatory role of the hydrogen storage in salt caverns is estimated as one means of the seasonal energy storage. It is assumed that all salt caves in Jiangsu are developed as the hydrogen storage containers to maximize the regulation potential of the hydrogen storage of Jiangsu salt caves in this paper. The main contributions of this paper are summarized as below.

1) a flexibility construction method is applied to the SCHSS-incorporated power system capacity planning model considering the high-resolution and long-term operational simulations

2) a one-year optimization simulation based on three different energy storage scenarios is conducted, and

comprehensive economic analysis is carried out

3) the long-term energy storage effect of SCHSS is verified, and the role of SCHSS in seasonal peak shaving is environmentally and economically proved

The remainder of this paper is organized as follows. Section 2 introduces the electro-hydrogen conversion and salt cavern hydrogen storage technology. Section 3 describes the power system planning model with SCHSS. In Section 4, case studies and discussions are presented and the conclusions are given in Section 5.

## 2. HYDROGEN PRODUCTION AND STORAGE

### *2.1 Electricity-to-hydrogen conversion*

Electricity-to-hydrogen conversion methods include electricity-to-hydrogen (P2H) and hydrogen-to-electricity (H2P), P2H generates hydrogen by electrolyzing water in the electrolytic tank, while H2P uses the fuel cells to generate electricity through an electrification process using the stored hydrogen.

There are many different types of electrolyzers used for electrolysis of hydrogen production, such as AEC (Alkaline Electrolyser), PEMEC (Proton Exchange Membrane Electrolyser), and SOEC (Solid Oxide Electrolyser). The characteristics of these three types of electrolyzers are shown in Table I.

***TABLE I**. Data of different electrolysis technologies in 2030*

| Electrolyzer type | Parameter description | | | |
|---|---|---|---|---|
| | Capital cost ($/kW) | O&M cost ($/kW/year) | Partial load range | System lifetime (years) |
| AEC | 604 | 30 | 20%-100% | 25 |
| PEMEC | 659 | 33 | 5%-100% | 15 |
| SOEC | 659 | 20 | 0-100% | 20 |

### *2.2 Hydrogen storage in salt caverns*

UHS uses an underground geological structure for large-scale hydrogen energy storage. When the power is excessive, the excess electricity is used to electrolyze water to produce hydrogen, and the hydrogen is injected into the underground geological structure such as aquifers, depleted oil and gas reservoirs, and salt caverns. The stored hydrogen is extracted during the peak of power demand, and power is generated through fuel cells to meet the power load demand. UHS is an effective way to solve large -capacity and long -term energy storage problems. Compared with other energy storage methods, UHS has the advantages of large scale, long storage period, low cost and high safety.

Salt caves are artificially built in the salt rock sedimentary layers. The advantage of salt cavern hydrogen storage is that the chemical reaction of rock salt and hydrogen is inert, and no impurities will be produced during the hydrogen storage process. In addition, salt caverns have good sealing performance. The diffusion loss of the hydrogen stored in salt caverns can be ignored, which provides great feasibility for storing pure hydrogen in salt caverns.

Fig.1 shows the power system structure including the SCHSS.

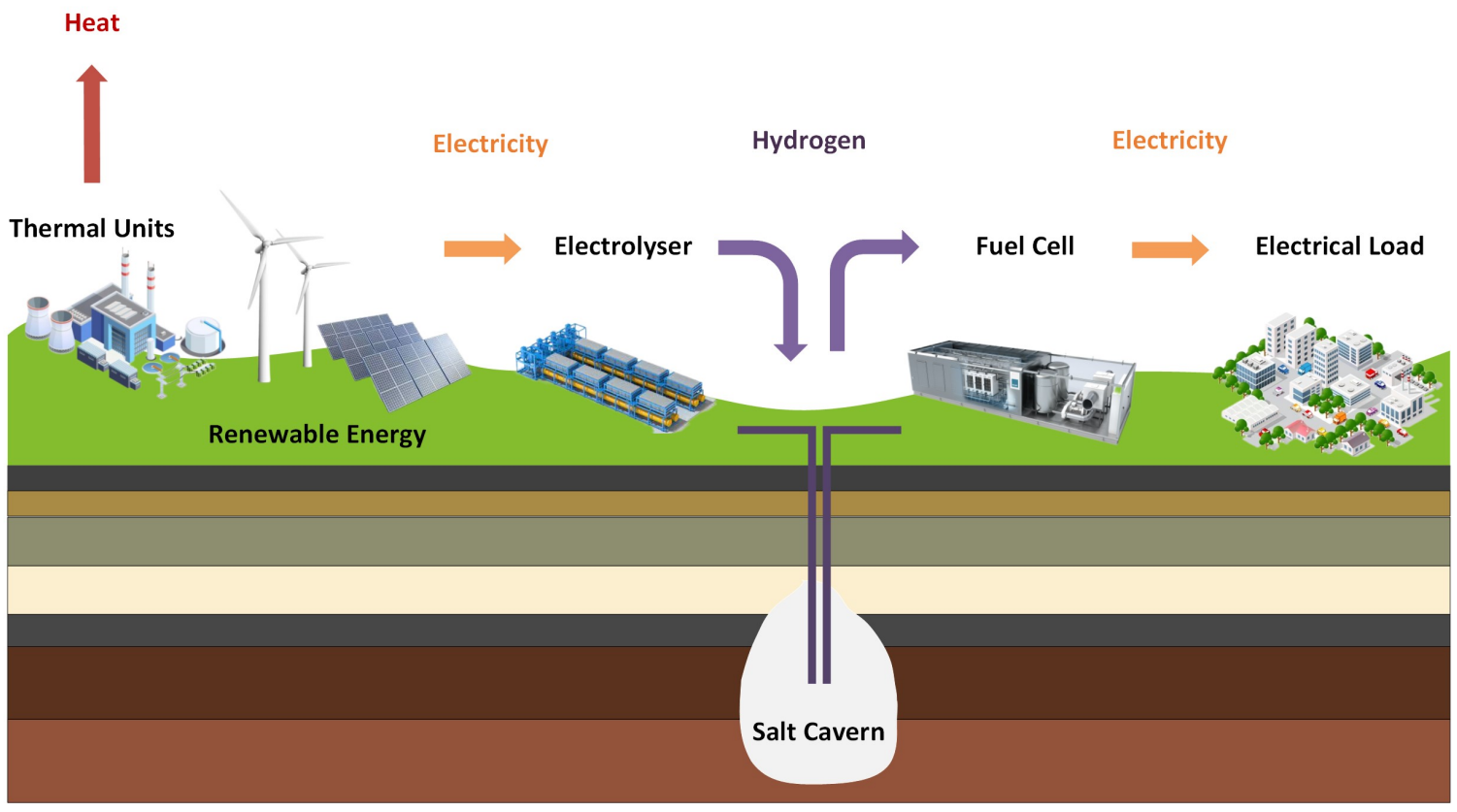

***Fig. 1*** *Power system including salt cavern hydrogen storage*

## 3. FORMULATION OF THE CAPACITY PLANNING MODEL

### *3.1 Improved Fast Unit Commitment*

1) Decision variables in the traditional model

Three binary variables are used to describe the operational status in the traditional model, which are $x_j(t)$, $s_j(t)$ and $u_j(t)$. These binary variables satisfy the following equation:

$$x_j(t) - x_j(t-1) = s_j(t) - u_j(t) \tag{1}$$

In Eq. (1), $x_j(t)$ is the on/off status of unit j. At time t, when the unit j is online, $x_j(t)$=1, When the unit j is offline, $x_j(t)$=0 at time t. $s_j(t)$ and $u_j(t)$ denote the startup and shutdown behavior of the unit respectively, indicating whether the unit j starts ($s_j(t)$=1) or shuts down ($u_j(t)$=0) at time t.

$p_j(t)$ is constrained by its operational range:

$$\overline{\delta_j} \cdot p_j^{max} \cdot x_i(t) \le p_j(t) \le \underline{\delta_j} \cdot p_j^{max} \cdot x_j(t) \tag{2}$$

where $p_j^{max}$ indicates the nameplate capacity of unit j, $\overline{\delta_j}$ and $\underline{\delta_j}$ represent the ratios of the minimum and maximum power output of unit j to its nameplate capacity.

For a group of units, three variables are used to model the overall status of units:

$$\widehat{P_t^i} = \sum_{j=1}^{J} \left( p_j^{max} \cdot x_j(t) \right) \tag{3}$$

$$\widehat{S_t^i} = \sum_{j=1}^{J} \left( p_j^{max} \cdot s_j(t) \right) \tag{4}$$

$$\widehat{U_t^i} = \sum_{j=1}^{J} \left( p_j^{max} \cdot u_j(t) \right) \tag{5}$$

where J is the number of units in the group $i$. $\widehat{P_t^i}$ , $\widehat{S_t^i}$ and $\widehat{U_t^i}$ respectively indicate the online capacity, startup capacity, and shutdown capacity of the group $i$ at time t.

As indicated by Eq. (3)-(5), the possible values of $\widehat{P_t^i}$ , $\widehat{S_t^i}$ and $\widehat{U_t^i}$ would be determined by combinations of $x_j(t)$ , $s_j(t)$ and $u_j(t)$ . Therefore, $\widehat{P_t^i}$ , $\widehat{S_t^i}$ and $\widehat{U_t^i}$ would be discrete sets of variables.

2) Fast Unit Commitment method

This paper uses three continuous variables ( $\overline{p}_t^i$ , $s_t^i$ and $u_t^i$ ) to approximate the discrete variables in Eq. (3)-(5).

$$0 \le \overline{p}_t^i, s_t^i, u_t^i \le S_i \tag{6}$$

$$S_i = \sum_{i=1}^{J} p_j^{max} \tag{7}$$

where $S_i$ represents the total capacity of group $i$. The relationship between this set of continuous variables is similar to Eq. (1):

$$\overline{p}_t^i - \overline{p}_{t-1}^i = s_t^i - u_t^i \tag{8}$$

This method transforms the integer programming problem into a linear programming problem, reduces the number of decision variables, and improves computational efficiency. The objective function and constraints are formulated based on the above continuous decision variables, which will be described later.

3.2 Objective Function

The objective of the comprehensive power system capacity planning model considering SCHSS is to minimize the operation cost of the electric hydrogen system:

$$min\ W_a + W_f + W_v + W_h \tag{9}$$

In the above formulas, $\mathrm{W_a}$ indicates the investment cost of newly-built units; $\mathrm{W_f}$ is the operation and maintain cost of thermal units and variable renewable energy units; $W_v$ indicates both the start-up cost and the fuel cost of thermal units; $W_h$ represents the cost of SCHSS.

$W_f$, $W_v$ and $W_h$ are given as follows:

$$W_a = \sum_{i=1}^{N} \left( a^i \cdot I_{\mathrm{new}}^i + a^w \cdot I_{\mathrm{new}}^w + a^s \cdot I_{\mathrm{new}}^s \right) \tag{10}$$

$$W_f = \sum_{i=1}^{N} \left( f^i \cdot I^i + f^w \cdot I^w + f^s \cdot I^s \right) \tag{11}$$

$$W_v = \sum_{i=1}^{N}\sum_{t=1}^{T}\left(c^i \cdot p_t^i \cdot \Delta t\right) + \sum_{i=1}^{N}\sum_{t=1}^{T}\left(SD^i \cdot s_t^i \cdot \Delta t\right) \tag{12}$$

$$W_h = a_{\text{salt}} \cdot I_{\text{salt}} + \sum_{t=1}^{T}\left(f_{\text{P2H}} \cdot P_{\text{P2H,t}} \cdot \Delta t + f_{\text{H2P}} \cdot H_{\text{H2P,t}} \cdot \Delta t\right) \tag{13}$$

*3.3 Constraints*

1) Conventional Power System Operational Constraints

A. Hourly Output Power Constraints

$$0 \le p_t^i \le \overline{p}_t^i \le I^i \tag{14}$$

$$0 \le p_t^w \le \alpha_t \cdot I^w \tag{15}$$

$$0 \le p_t^s \le \beta_t \cdot I^s \tag{16}$$

In this formula, $\overline{p}_t^i$ is the continuous variable obtained using the fast unit combination method.

B. Ramping Constraints

$$p_t^i \le \overline{\mu}_i \cdot \left(\overline{p}_t^i - s_t^i - u_{t+1}^i\right) + VS^i \cdot s_t^i + VD^i \cdot u_{t+1}^i \tag{17}$$

$$p_t^i - p_{t-1}^i \le RU^i \cdot \left(\overline{p}_t^i - s_t^i\right) + VS^i \cdot s_t^i - \underline{\mu}_i \cdot u_t^i \tag{18}$$

$$p_{t-1}^i - p_t^i \le RD^i \cdot \left(\overline{p}_t^i - s_t^i\right) - \underline{\mu}_i \cdot s_t^i + VD^i \cdot u_t^i \tag{19}$$

C. Minimum on/off Time Constraints

$$0 \le u_{t+1}^i \le \overline{p}_t^i - \sum_{\tau=0}^{t-1} s_{t-\tau}^i, t = 1 \cdots UT^i - 1 \tag{20}$$

$$0 \le u_{t+1}^i \le \overline{p}_t^i - \sum_{\tau=0}^{UT_k^i - 2} s_{t-\tau}^i, t = UT^i \cdots T - 1 \tag{21}$$

$$0 \le s_{t+1}^i \le I^i - \overline{p}_t^i - \sum_{\tau=0}^{t-1} u_{t-\tau}^i, t = 1 \cdots DT^i - 1 \tag{22}$$

$$0 \le s_{t+1}^i \le I^i - \overline{p}_t^i - \sum_{\tau=0}^{DT_k^i - 2} u_{t-\tau}^i, t = DT^i \cdots T - 1 \tag{23}$$

Eq. (20)-(23) states that a fixed amount of time must elapse after a thermal unit has been switched on/off before it can be switched off/on.

D. Thermoelectric Ratio Constraint

$$p_t^e = TE \cdot p_t^h \tag{24}$$

There is a fixed ratio between the power generation and the heat production of CHP units, $TE$ denotes the thermoelectric ratio of CHP units.

E. Energy Balance Constraints

$$p_t^e = HM_t \tag{25}$$

$$\sum_{i=1}^{N} P_t^i + p_t^w + p_t^s + p_{\mathrm{H2P},t} - p_{\mathrm{P2H},t} = DM_t \tag{26}$$

According to Eq. (26), all types of generator units and SCHSS cooperate to meet the power load demand.

2) Hydrogen Storage System Constraints

A. Electricity-hydrogen conversion constraint

$$0 \le p_{\mathrm{P2H},t} \le p_{\mathrm{P2H},max} \tag{27}$$

$$0 \le H_{\mathrm{H2P},t} \le H_{\mathrm{H2P},max} \tag{28}$$

$$H_{\mathrm{P2H},t} = p_{\mathrm{P2H},t}\eta_{\mathrm{P2H}} \tag{29}$$

$$p_{\mathrm{H2P},t} = H_{\mathrm{H2P},t}\eta_{\mathrm{H2P}} \tag{30}$$

According to Eq. (26), the maximum electric power of electrolytic hydrogen production units and the maximum hydrogen consumption of fuel cells are limited. The degree of energy loss of the electrical-hydrogen conversion process is affected by the hydrogen production efficiency of electrolysis units and the power generation efficiency of fuel cells.

B. Hydrogen storage constraint

$$\begin{cases} H_{H,\mathrm{in},t}\eta_{H,\mathrm{in}} - \dfrac{H_{H,\mathrm{out},t}}{\eta_{H,\mathrm{out}}} - S_{H,t}L_H = S_{H,t} - S_{H,\mathrm{ini}}, t = 1 \\ H_{H,\mathrm{in},t}\eta_{H,\mathrm{in}} - \dfrac{H_{H,\mathrm{out},t}}{\eta_{H,\mathrm{out}}} - S_{H,t}L_H = S_{H,t} - S_{H,t-1}, t > 1 \end{cases} \tag{31}$$

$$0 \le S_{H,t} \le S_{H,max} \tag{32}$$

Eq. (31) represents the hydrogen storage balance constraint of SCHSS, and Eq. (32) represents the hydrogen storage amount constraint of salt caverns.

## 4. CASE STUDY

The power consumption and the scale of salt caverns of Jiangsu Province rank high in China, making it the best province to test the effect of SCHSS. The electricity consumption in Jiangsu in 2021 exceeded 700 billion kWh, ranking third among all the provinces in China. The electricity load in Jiangsu Province has obvious seasonal characteristics, with the peak periods in summer and winter, and the low periods in spring and autumn. The peak of the summer electricity consumption in Jiangsu Province is caused by using air conditioners. In 2022, the maximum load power of Jiangsu in summer reached 103.6 million kW. The peak of the electricity consumption in Jiangsu Province in winter is due to the use of electric heating. The data shows that the maximum load power of Jiangsu power grid in winter reached 117 million kW in 2021. In winter and summer, the peak-to-valley difference in the electricity consumption in Jiangsu Province can reach 35%. In addition, Jiangsu Province is rich in salt mine resources and the development potential of salt caverns is huge. There are three rock salt mines with their reserves greater than 10 billion tons in Jiangsu Province, accounting for 37.5% of the total of China [49].

Fig.2 shows the distribution of the salt mines in China. It can be seen from Fig. 2 that there are a large number of salt mines in Jiangsu Province, which makes it have great potential for the hydrogen storage in salt caverns. Therefore, it is of great significance to study the application potential of the salt cavern hydrogen storage in Jiangsu Province. This paper is based on the projection to 2050 with hourly operation simulation for the whole year.

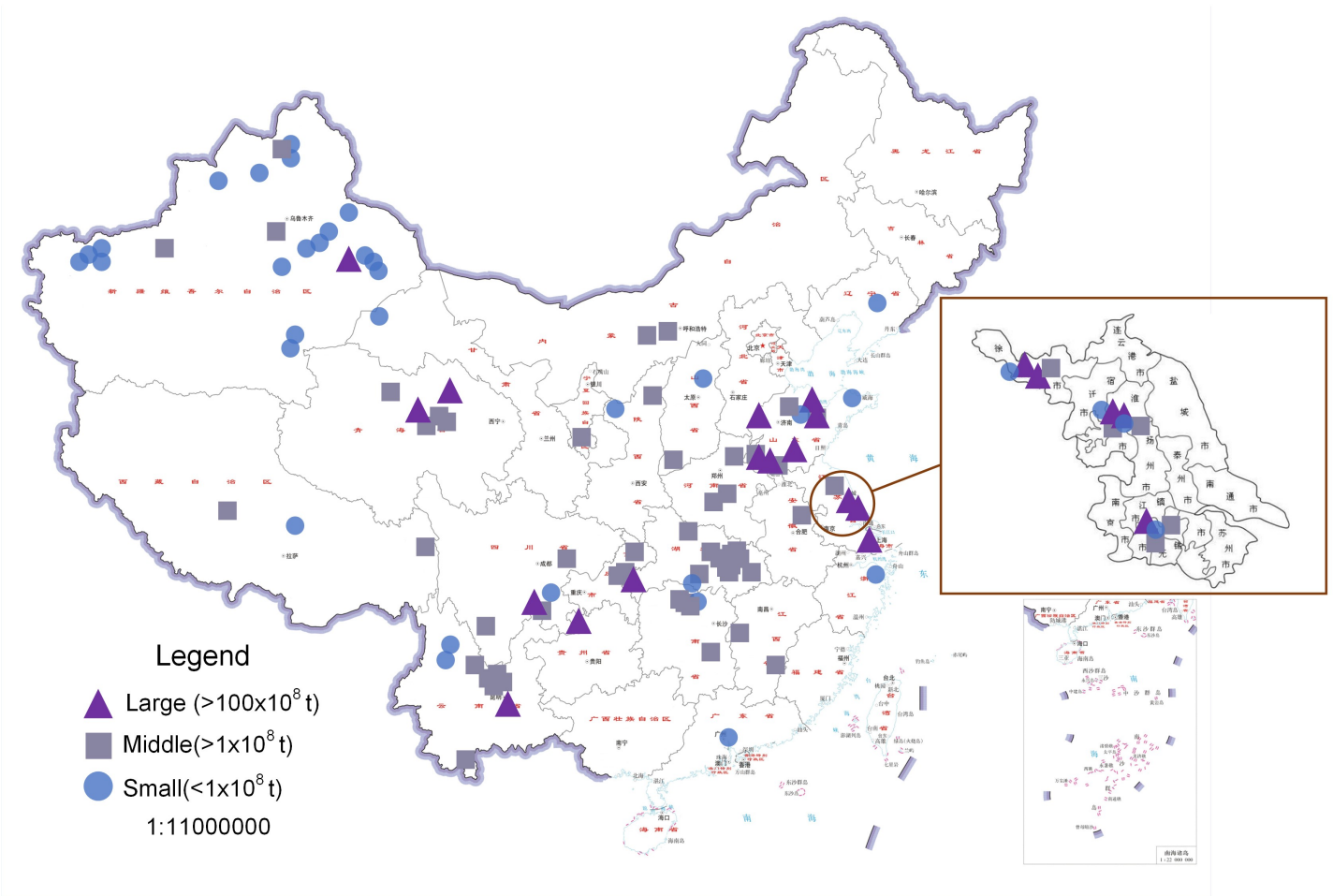

***Fig. 2*** *Distribution of salt mines in China*

The installed capacity ratio of different generation units in Jiangsu Province in 2021 is shown in Fig.1. The relevant unit capacity and the operation constraints data are presented in Table II.

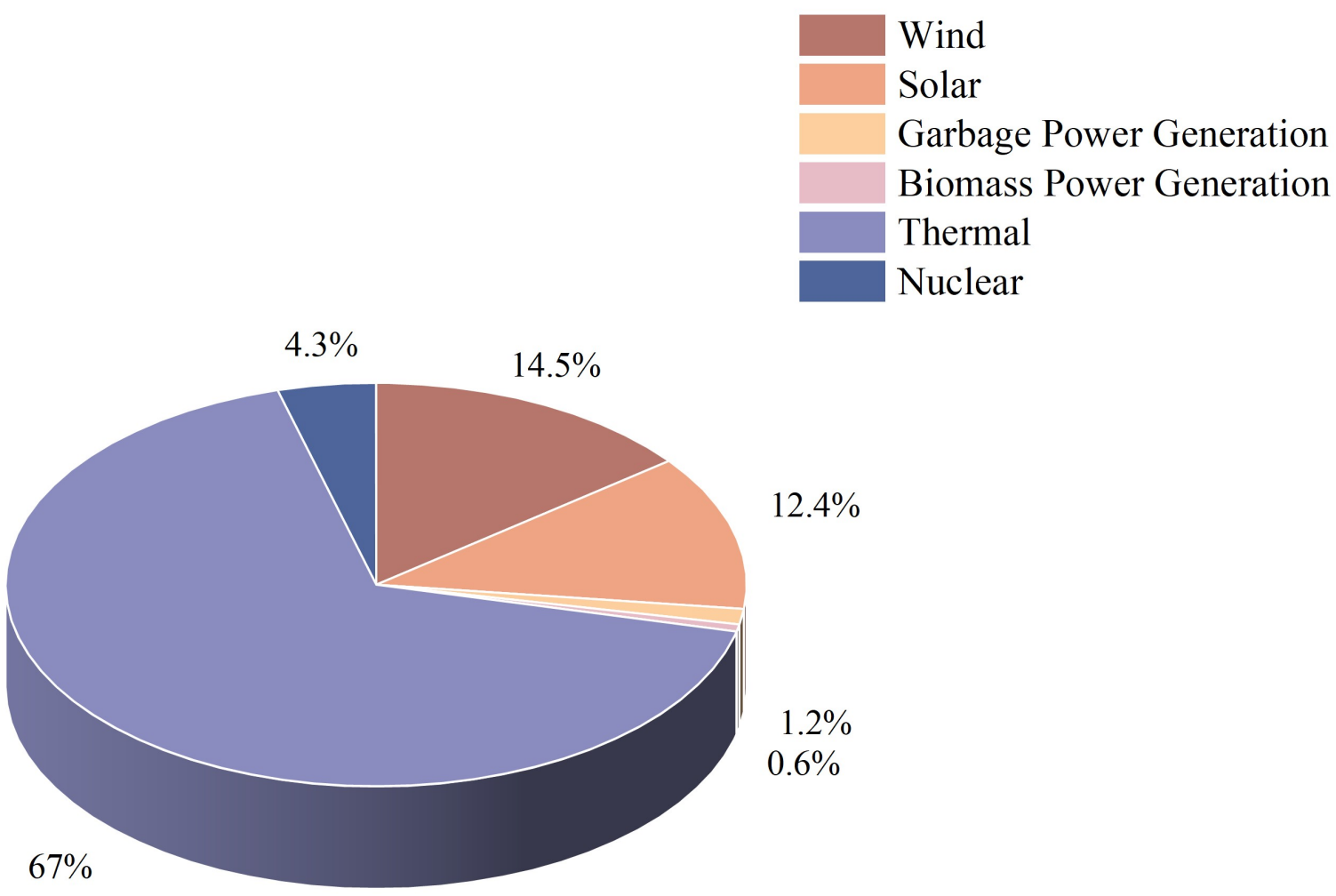

*Fig.3 Installed capacity structure of Jiangsu Province*

***TABLE II***. *Installed capacity and operation characteristics of thermal power units in Jiangsu Province*

| Parameter information | Coal | CHP | Gas |
|---|---|---|---|
| Capital cost ($/kW) | 621 | 621 | 524 |
| O&M cost (% of capital cost) | 2.1 | 2.1 | 2.6 |
| Start-up cost ($/MW) | 147 | 147 | 88 |
| Minimum up time (hours) | 8 | 8 | 1 |

In this paper, 3 energy storage scenarios are considered. For each scenario, different variable renewable energy penetration rates are simulated. The three simulation scenarios are indicated in Table III.

***TABLE III***. *Simulation scenario introduction*

| Scenarios | Energy storage technology |
|---|---|
| BAU | Traditional electrochemical energy storage |
| HSS | Hydrogen storage above ground (take high-pressure hydrogen storage tank as an example) |
| SCHSS | Hydrogen storage in underground salt caverns |

A one-year simulation based on Yalmip is carried out on MATLAB, and the results obtained by Gurobi are as follows.

### *4.1 A. Analysis of simulation results in different scenarios*

1) BAU

Fig. 4 depicts the hourly power output of different units at 20%, 40%, 60% and 80% penetration in the BAU scenario. With the increase of the penetration rate of variable renewable energies, the outputs of wind energy and solar energy have increased significantly.

When the penetration rate of variable renewable energies is low, the chemical energy storage has little effect, which is due to the high cost of the chemical energy storage. The other reason is that thermal unit has high flexibility, when variable renewable energy penetration is below 40%, the capacity of thermal units is sufficient to meet the peak-shaving requirements, so there is no need to introduce energy storage to increase the flexibility of the power system.

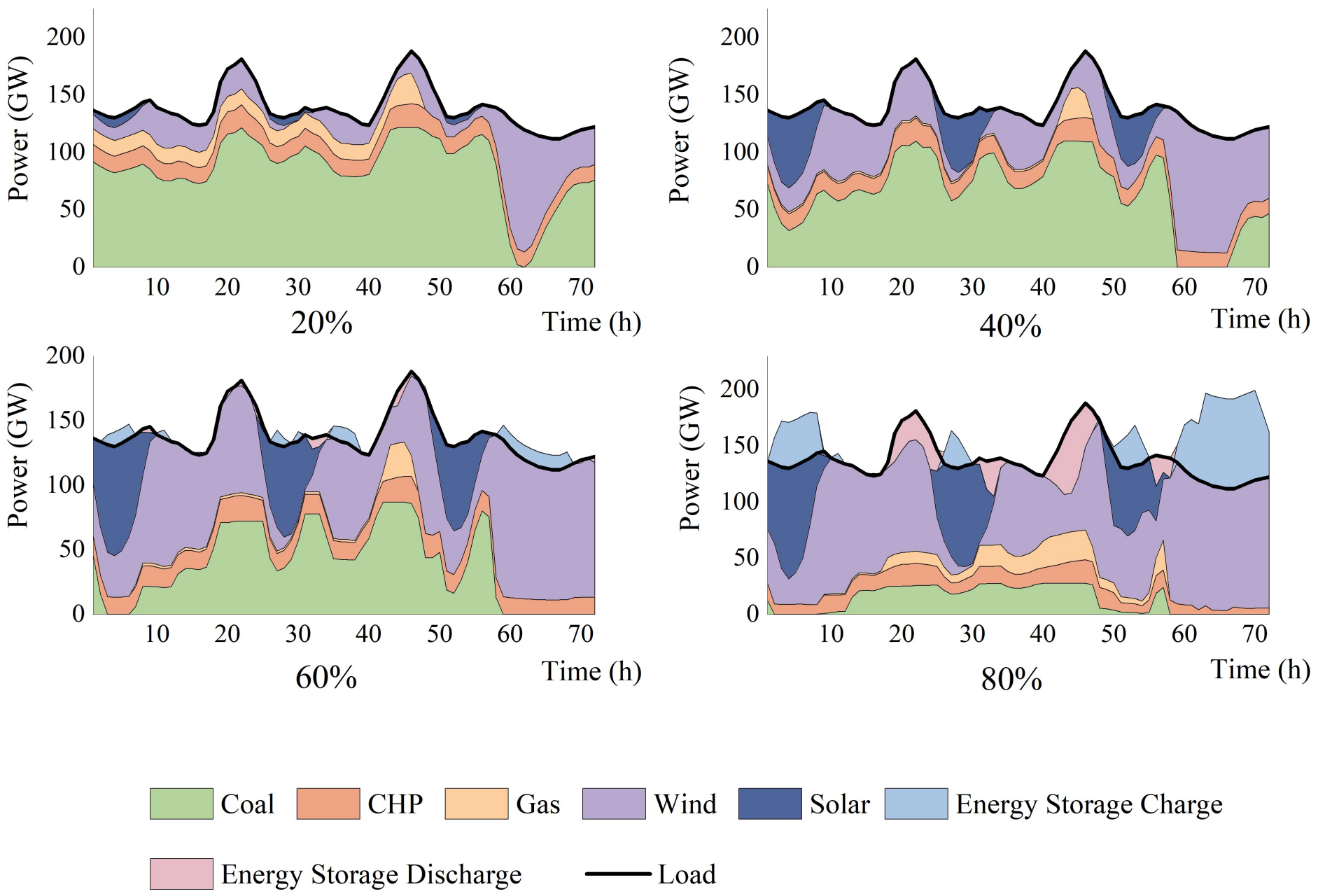

***Fig.4*** *Hourly power output of different units at different penetration rates (BAU scenario)*

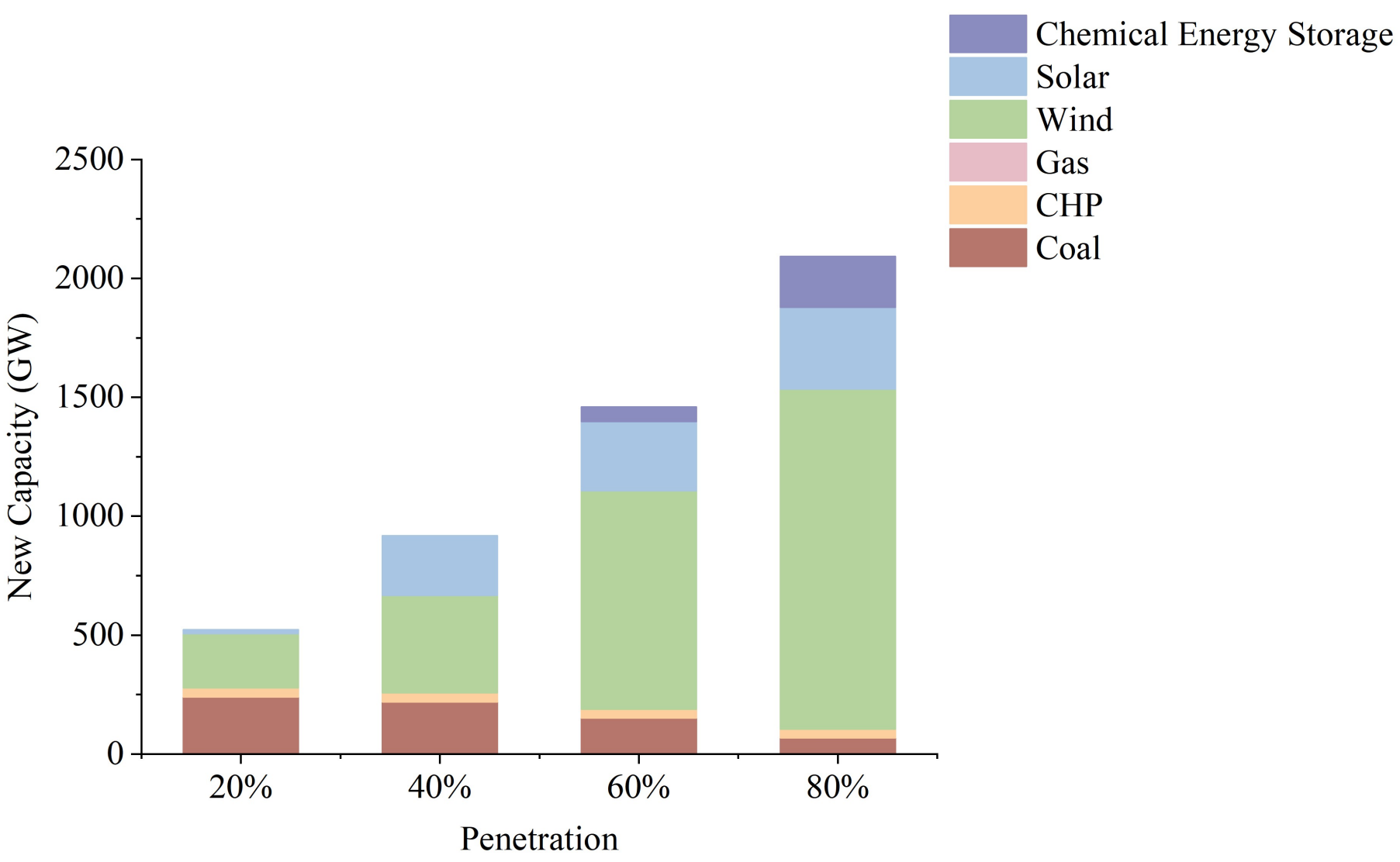

***Fig.5*** *Composition of new units at different penetration rates (BAU scenario)*

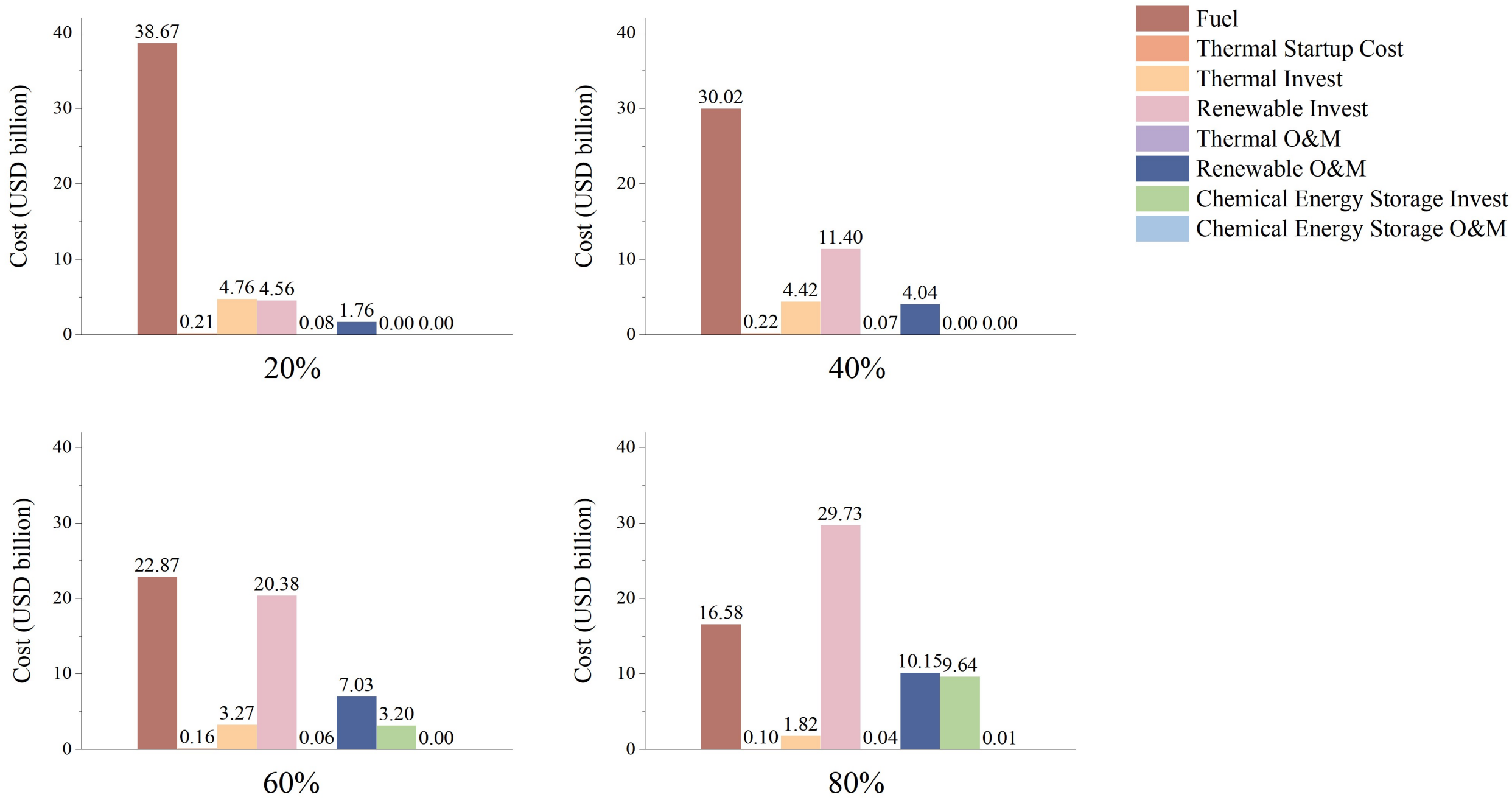

***Fig.6*** *Distribution of power system operation cost at different penetration rates (BAU scenario)*

The sizes of the newly-built capacities under different penetration rates in the BAU scenario are shown in Fig. 5. With the increase of the variable renewable energy penetration, the newly-added capacity of thermal power units has decreased significantly, while the newly-added capacity of variable renewable energies has increased significantly. Among them, the newly-built capacity of wind power has the most obvious increase, reaching more than half of the total newly-built capacity. Traditional chemical energy storage has a larger newly-added capacity when the variable renewable energy penetration is higher, which is because the proportion of thermal power capacity is low at this time, and the flexibility of the power system is insufficient. The energy storage is required to provide the peak-shaving and valley-filling functions.

Fig. 6 presents the distribution of the power system operation costs under different penetration rates in the BAU scenario. With the increase of variable renewable energy penetration, the fuel costs of thermal power units (coal, natural gas, etc.) gradually decrease, and the investment and operation and maintenance costs of variable renewable energy units gradually increase. When the variable renewable energy penetration is low, the cost mainly comes from thermal power units. When the variable renewable energy penetration is high, the cost mainly comes from the variable renewable energy units, while the fuel cost of thermal power units still accounts for a large proportion.

2) HSS

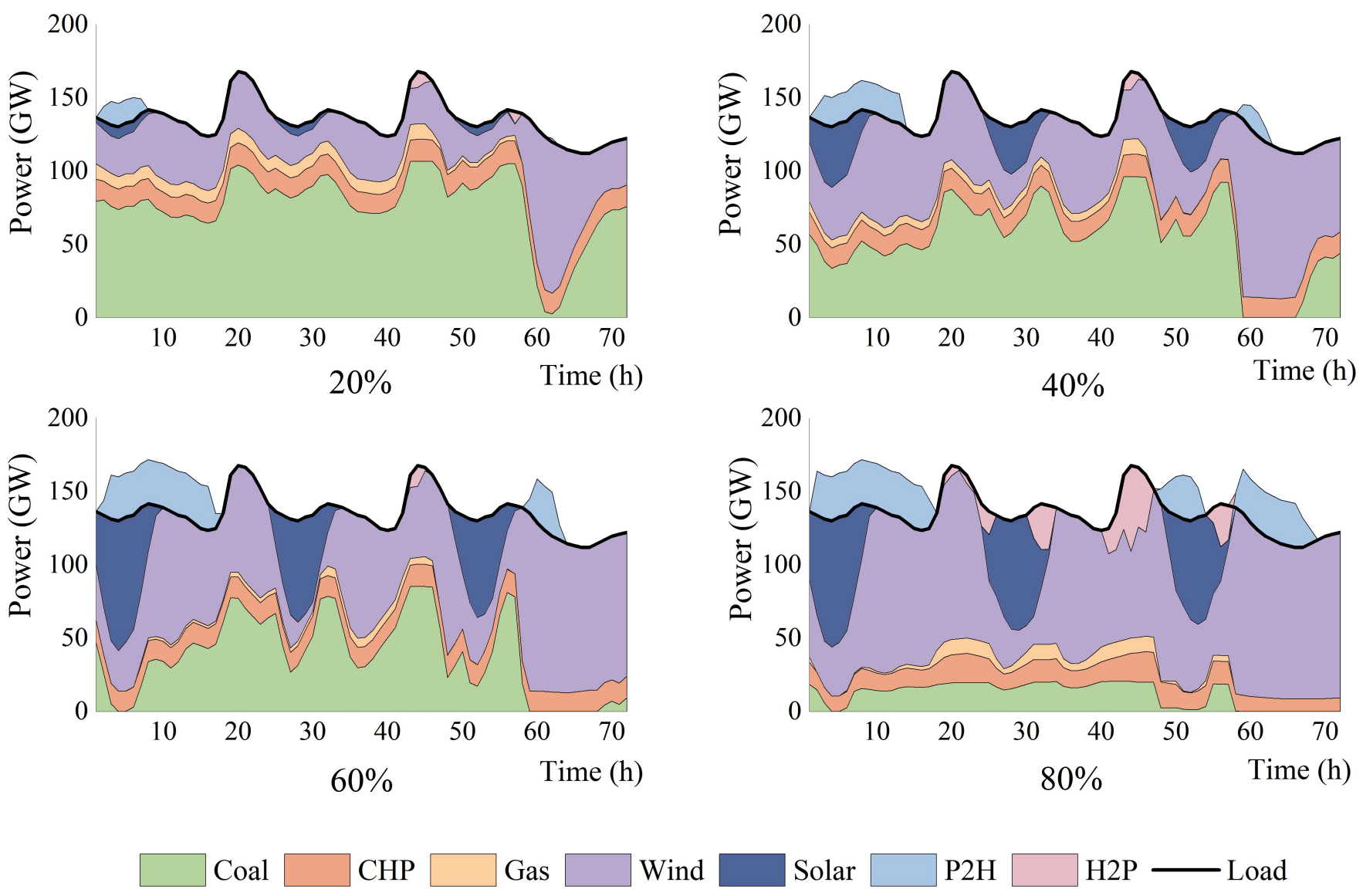

***Fig.7*** *Hourly power output of different units at different penetration rates (HSS scenario)*

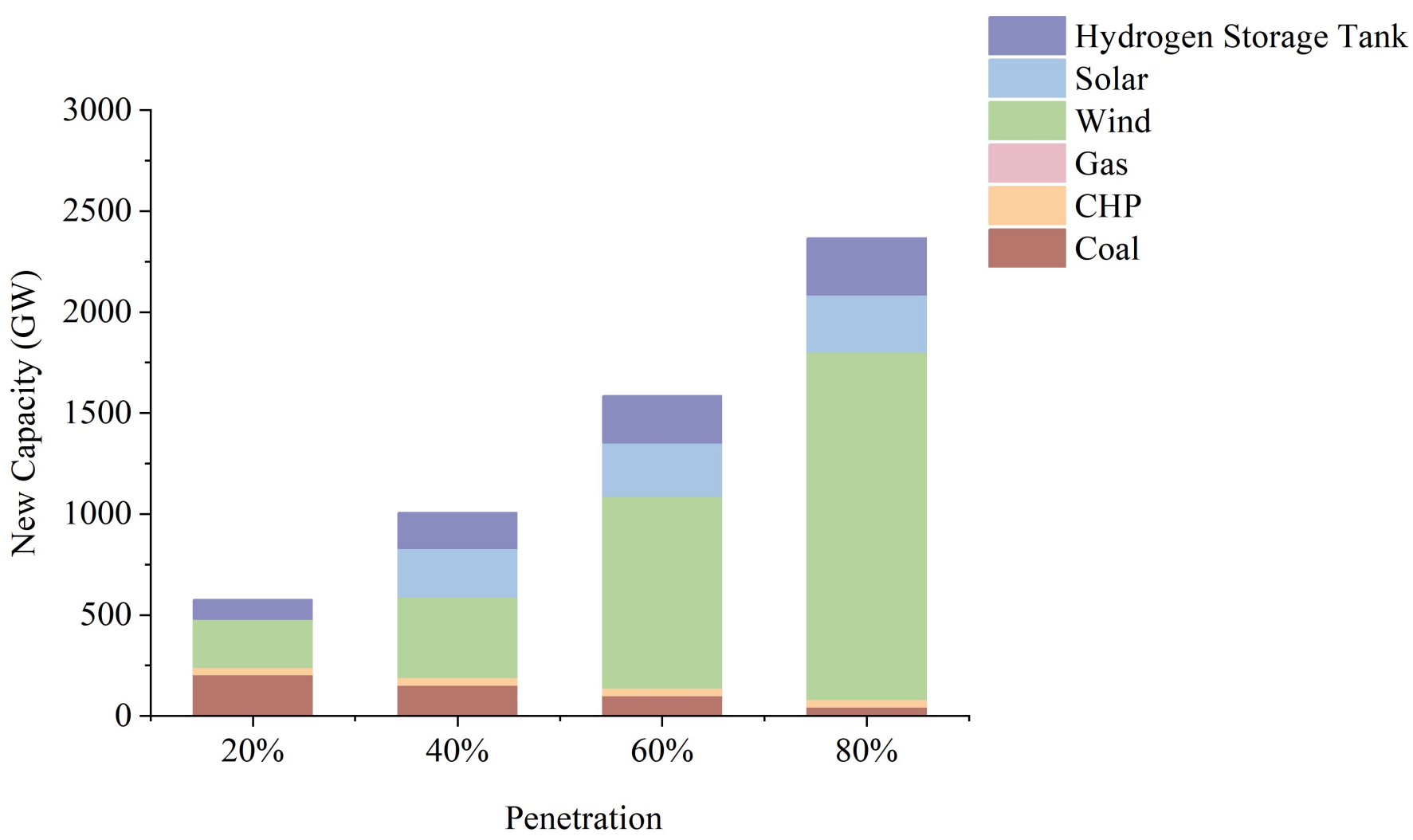

***Fig.8*** *Composition of new units at different penetration rates (HSS scenario)*

Fig.7 demonstrates the hourly power output of thermal power and variable renewable energy in the HSS scenario. Different from the BAU scenario, when the proportion of the variable renewable energy is low, the hydrogen storage tank can also play a role in peak regulation. When the variable renewable energy penetration increased to 80%, the charging/discharging power of the hydrogen storage tank does not increase significantly. This is due to the limitation of the maximum hydrogen injection and extraction flow rate of the hydrogen storage tank.

Fig.8 presents the size of the newly-built capacity of various units under different penetration rates in the HSS scenario. Under different variable renewable energy penetration rates, the newly-added capacity of the hydrogen storage tanks is increased compared with the conventional electrochemical energy storage devices.

The distribution of the power system operation costs under different penetration rates in the HSS scenario is depicted in Fig. 9. Compared with the BAU scenario, the fuel cost of thermal power units and the cost of energy storage are reduced.

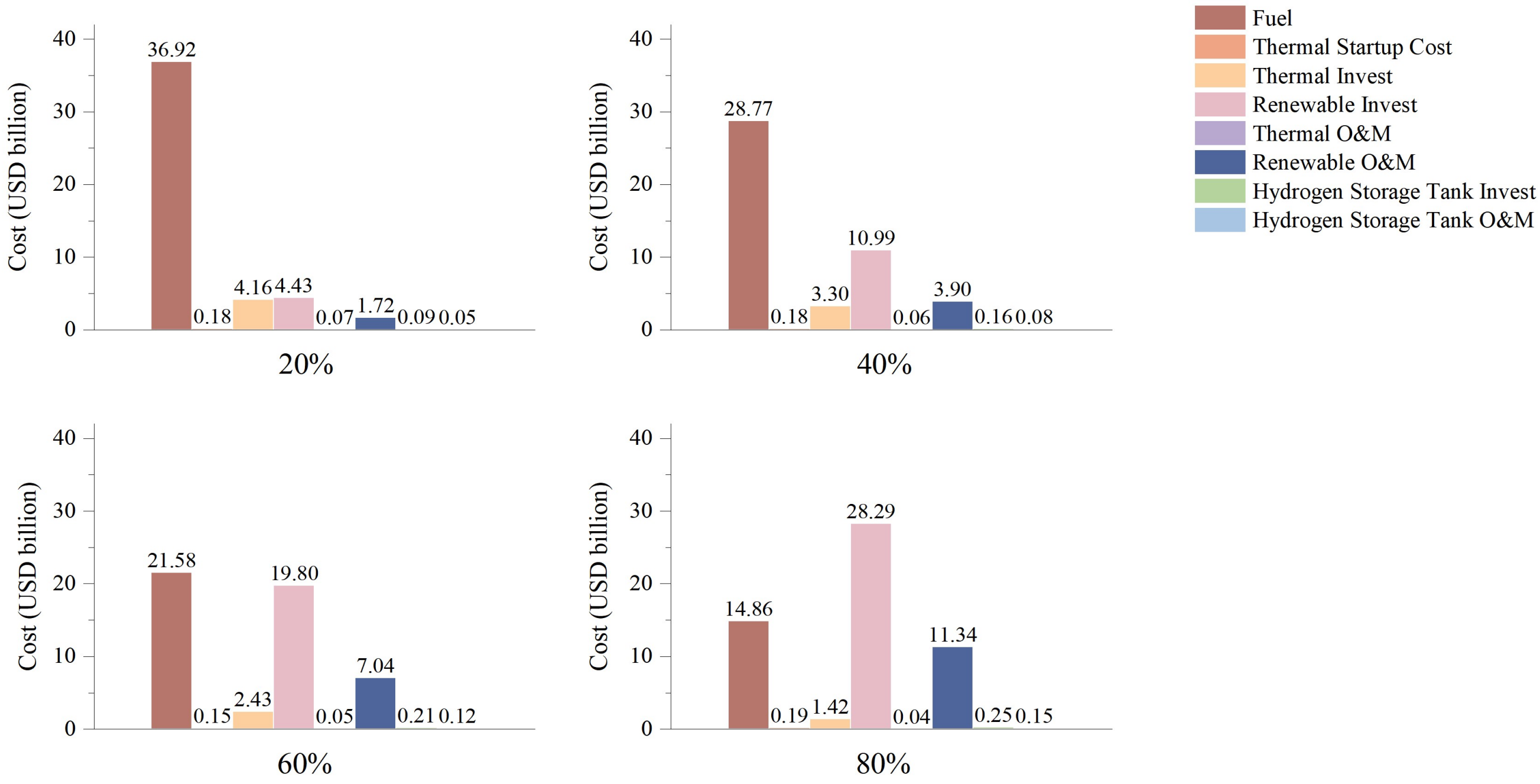

***Fig.9*** *Distribution of power system operation cost at different penetration rates (HSS scenario)*

3) SCHSS

Fig. 10 depicts the hourly power output of different units under different variable renewable energy penetration rates in the SCHSS scenario. It is visible that between the penetration of 20% and 80% of the variable renewable energy, SCHSS plays a crucial role in peak shaving and valley filling, which shows that the salt cavern hydrogen storage has better economy and flexibility.

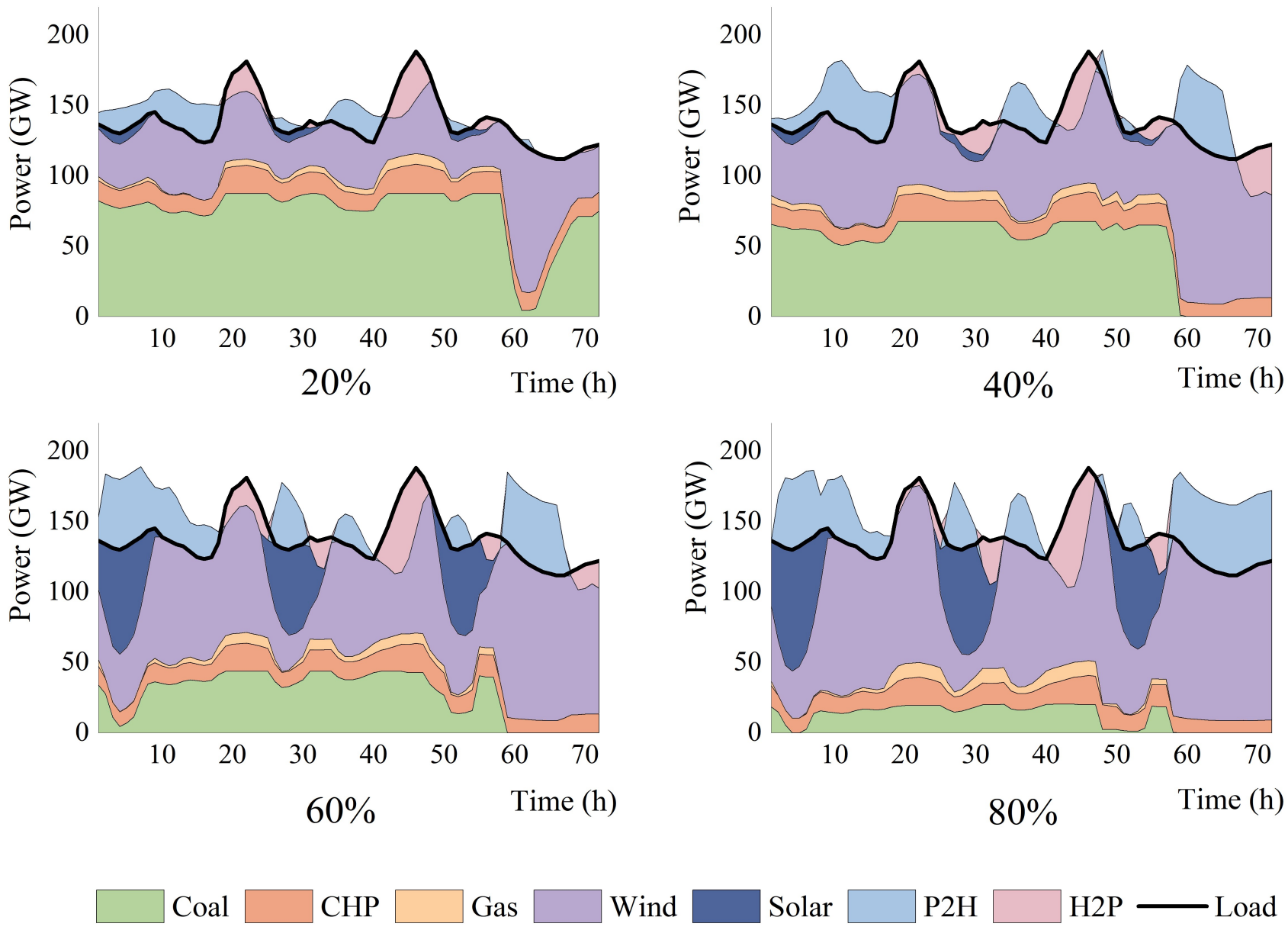

***Fig.10.*** *Hourly power output of different units at different penetration rates (SCHSS scenario)*

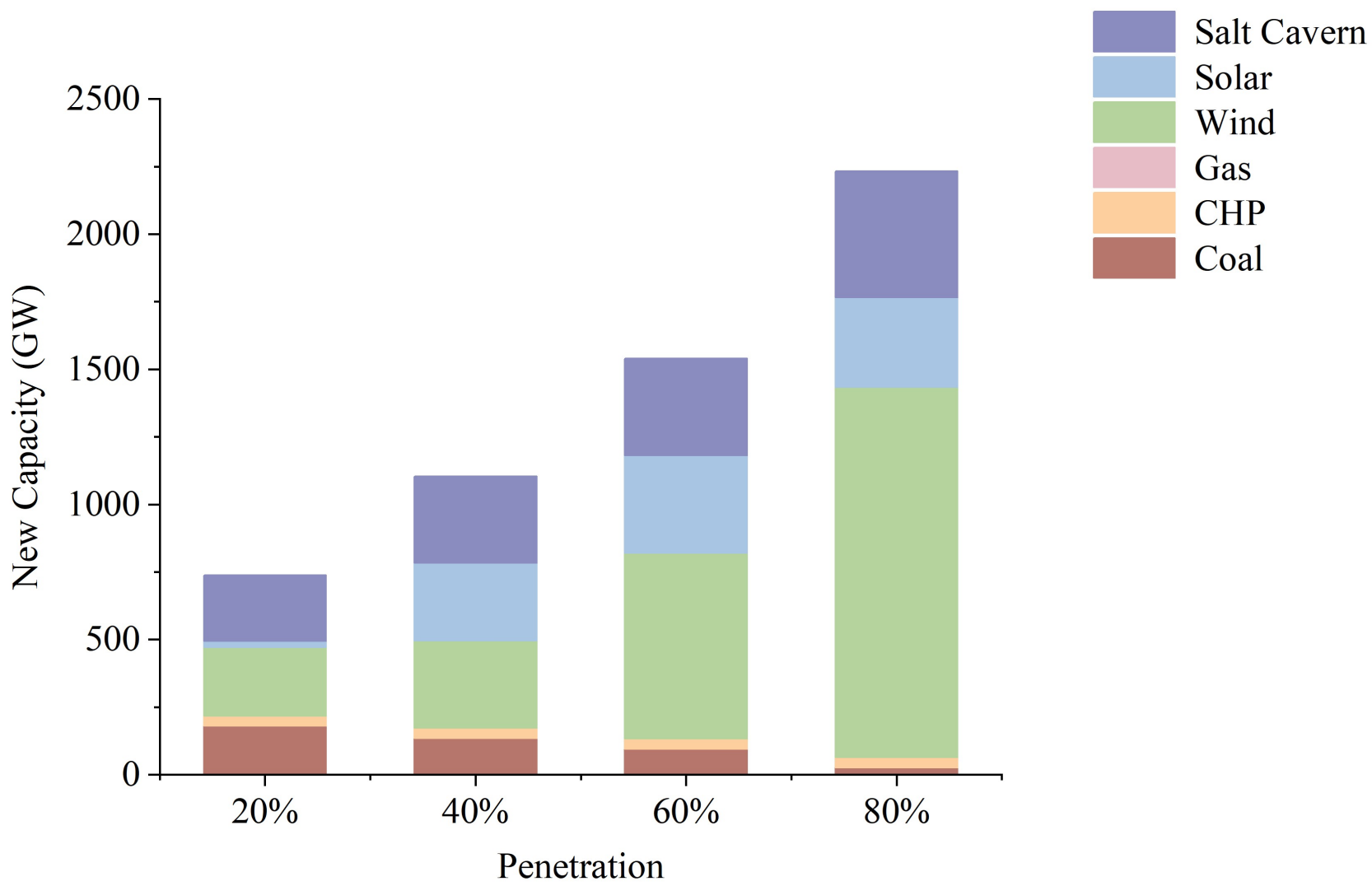

***Fig.11*** *Composition of new units at different penetration rates (SCHSS scenario)*

Fig. 11 shows the size of the newly-built capacity of various units at different penetration rates in the SCHSS scenario. The capacity of the newly-added salt cavern hydrogen storage is larger than the capacity of the newly-added electrochemical energy storage under the BAU scenario and the capacity of the newly-added hydrogen storage tank under the HSS scenario.

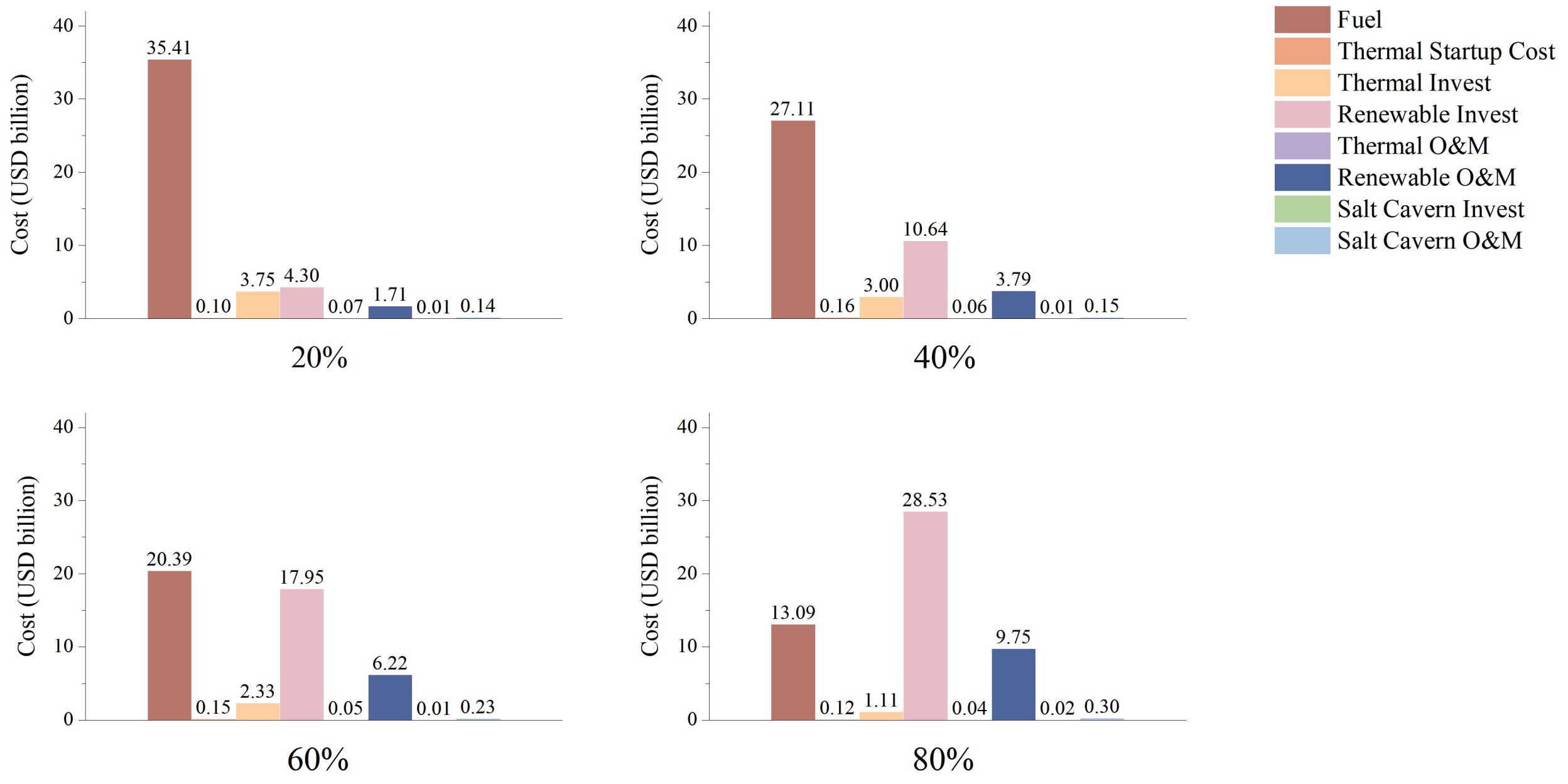

***Fig.12*** *Distribution of power system operation cost at different penetration rates (SCHSS scenario)*

Fig.12 demonstrates the distribution of the power system operation costs under different penetration rates in the SCHSS scenario. Compared with the previous two scenarios, although the new capacity of the salt cavern energy storage has been significantly increased, its investment cost is reduced to some extent, which shows that salt cavern hydrogen storage is the most economical energy storage method among the three scenarios. When variable renewable energy penetration is 80%, the overall cost of the power system under the SCHSS scenario is 22.2% lower than that of the BAU scenario, and 6.3% lower than that of the HSS scenario.

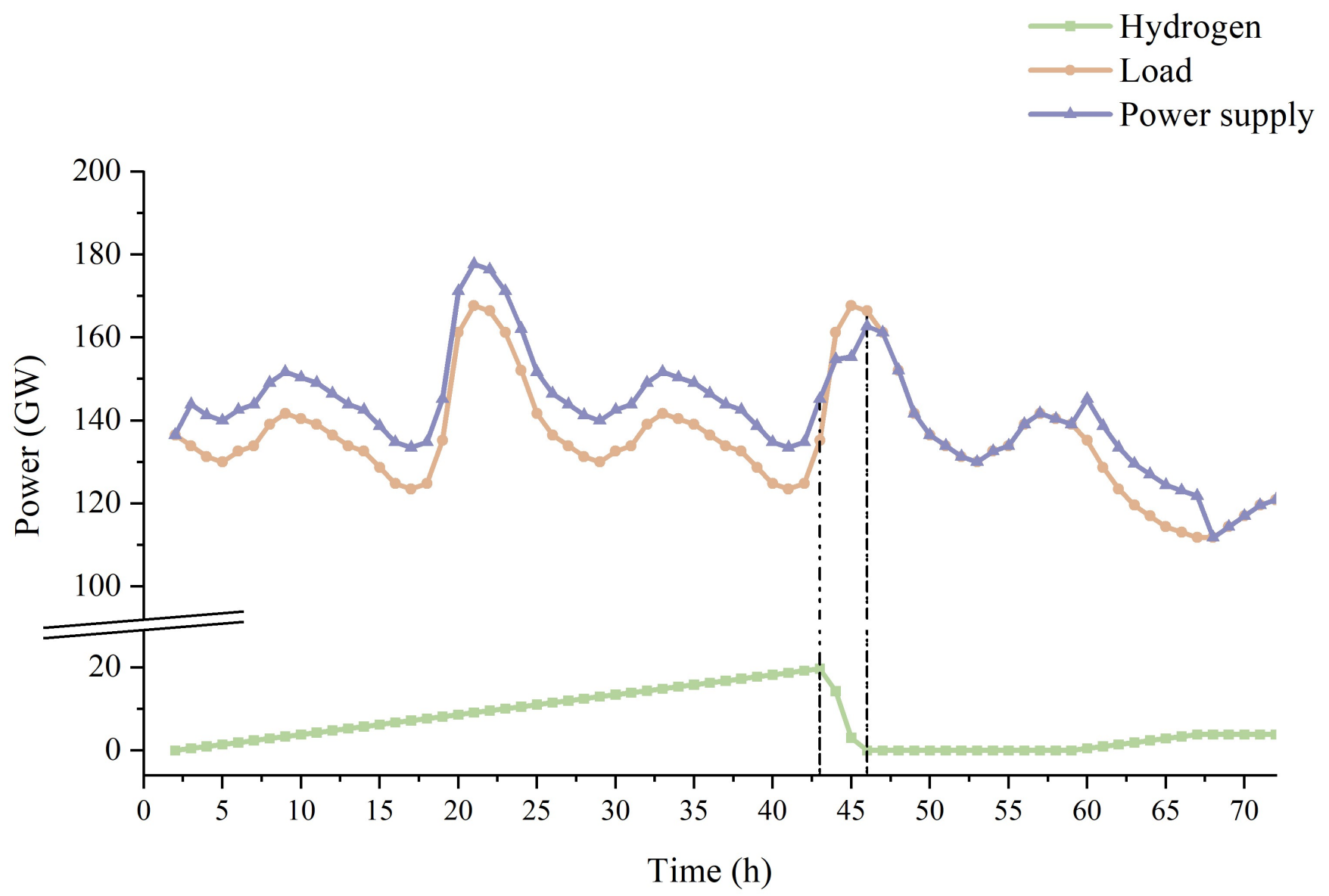

***Fig.13*** *SCHSS state of charge at 50% penetration*

Fig.13 shows the SCHSS state of charge at 50% of the variable renewable energy penetration. It is obvious that the salt cavern hydrogen storage can convert the remaining energy into the hydrogen stored in salt caverns when there is excess energy (the power generation is greater than the load), and the hydrogen can be extracted during the peak periods of the energy demand (the load is greater than the power generation). This shows that SCHSS can play an important role in peak shaving and valley filling.

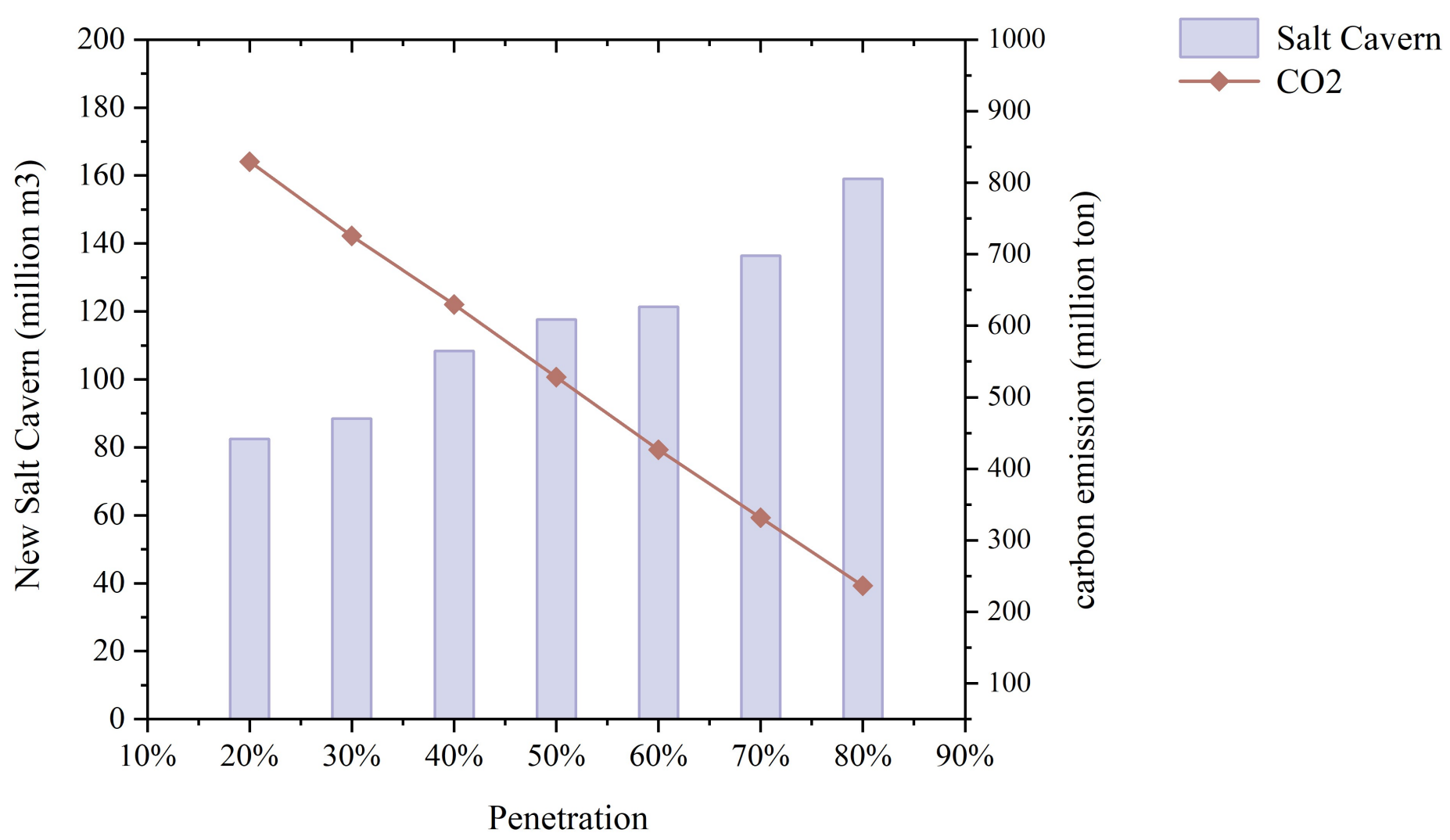

***Fig.14*** *Carbon emissions and capacity of newly-built salt caverns at different penetration rates*

Fig.14 indicates the carbon emissions of the power system and the newly-built capacity of salt caverns under different variable renewable energy penetration rates. It can be seen that with the increase of the variable renewable energy penetration rate, the carbon emission of the power system is significantly reduced, the newly-built capacity of salt caverns increases. When the variable renewable energy penetration increases from 20% to 80%, the CO2 emissions are reduced by 71.5%, and the newly-built capacity of SCHSS increases by 92.7%.

### *4.2 Comparison of three scenarios*

In order to better demonstrate the superiority of SCHSS, the simulation results of the three scenarios are compared.

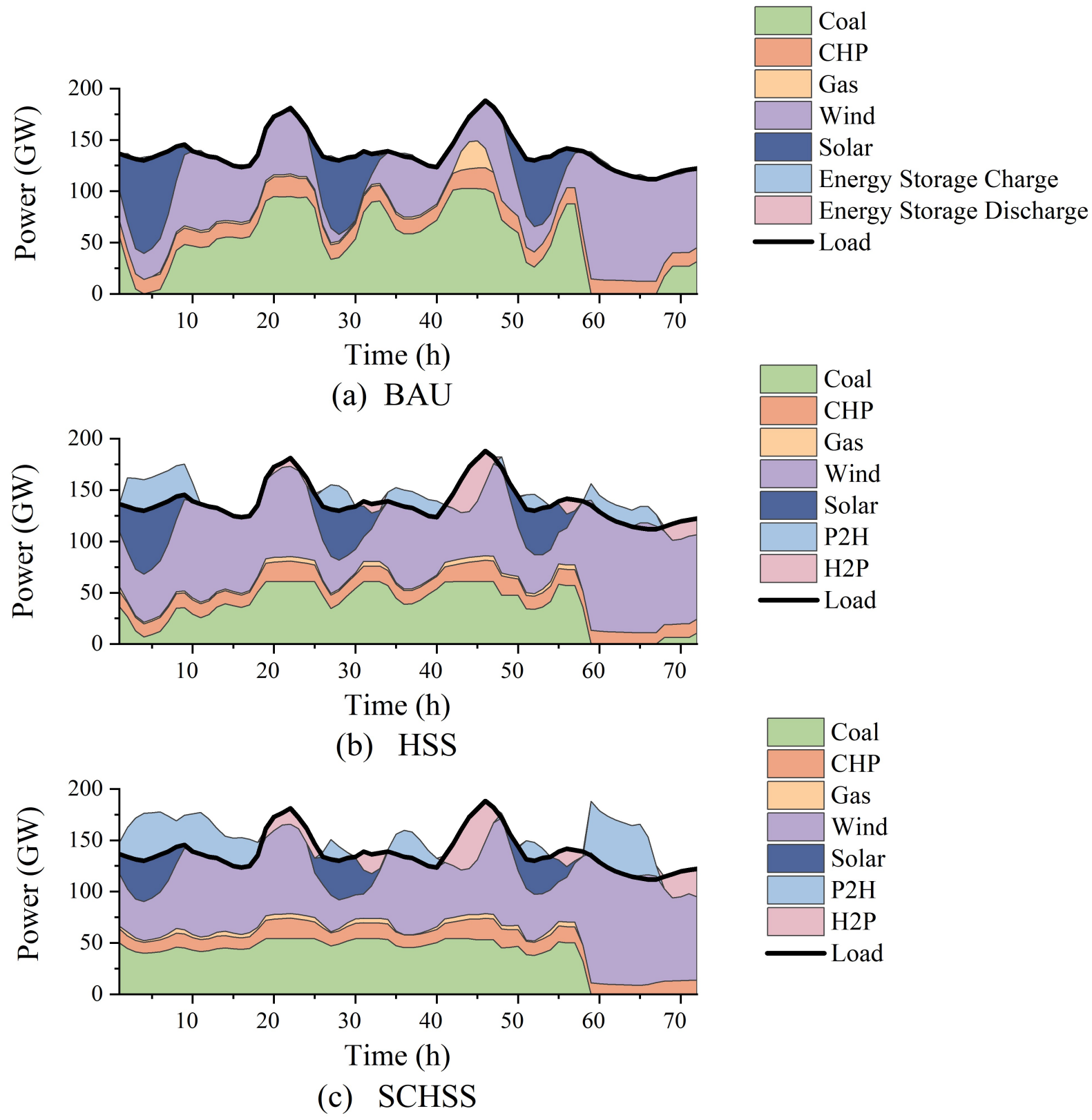

***Fig.15.*** *Hourly output of different units at 50% penetration*

Fig. 15 shows the hourly output of different units under different scenarios when the penetration rate is 50%. It can be found that the peak-shaving and valley-filling effects of the energy storage in the SCHSS scenario is the most obvious.

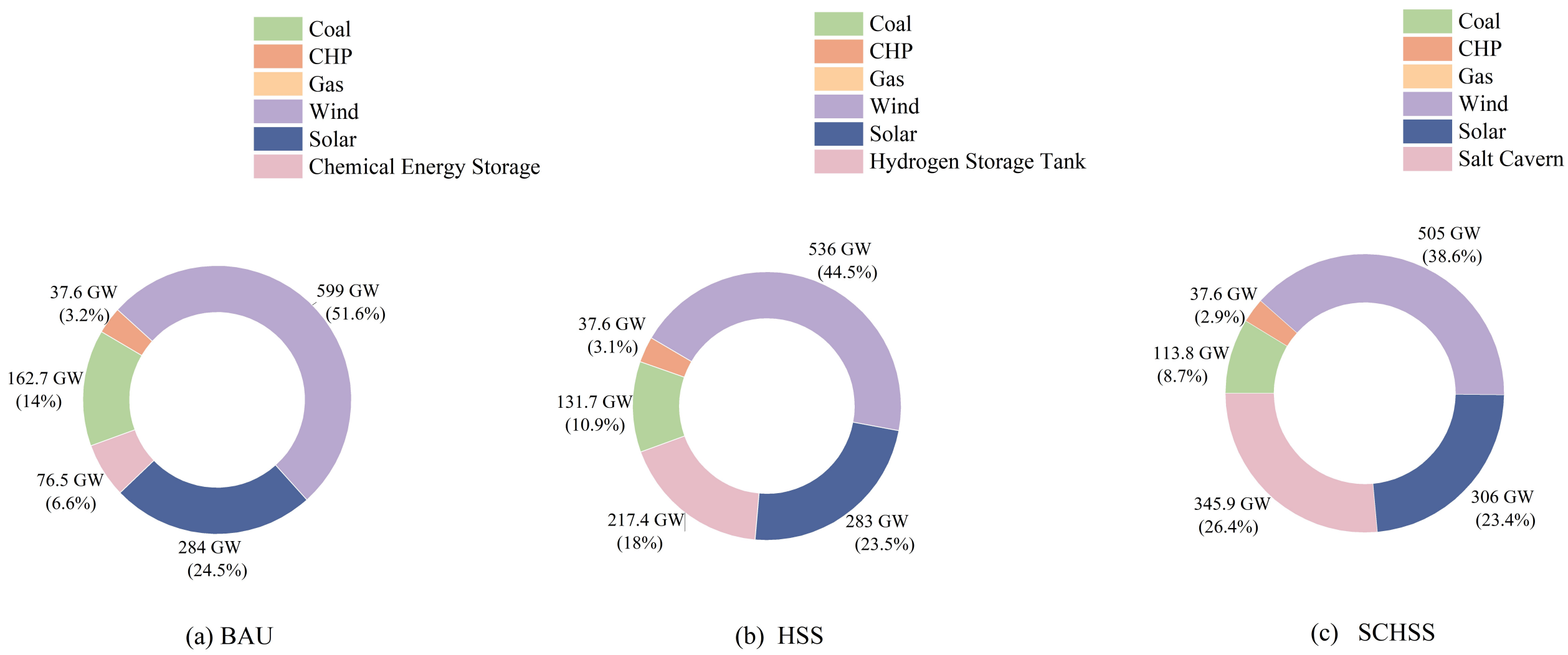

***Fig.16*** *Capacity distribution of new units at 50% penetration*

Fig. 16 depicts the composition of new units in three scenarios of BAU, HSS, and SCHSS with a penetration rate of 50%. It shows that the newly-built capacity of the energy storage in the SCHSS scenario is the largest, and the newly-built capacity of salt caverns is 4.52 times that of the traditional electrochemical energy storage and 1.59 times that of the hydrogen storage tanks. This is because the levelized cost of the SCHSS is the lowest among the three energy storage scenarios. In addition, the newly-built capacity of thermal units and variable renewable energy units under the SCHSS scenario is the lowest among the three scenarios. This is because energy storage can replace thermal units and play a role in improving system flexibility. The construction of a large number of energy storage can reduce the newly-built capacity and output of thermal units. Under a certain variable renewable energy penetration rate, newly-built thermal capacity decreases, and the newly-built capacity of renewable energy can also be reduced accordingly.

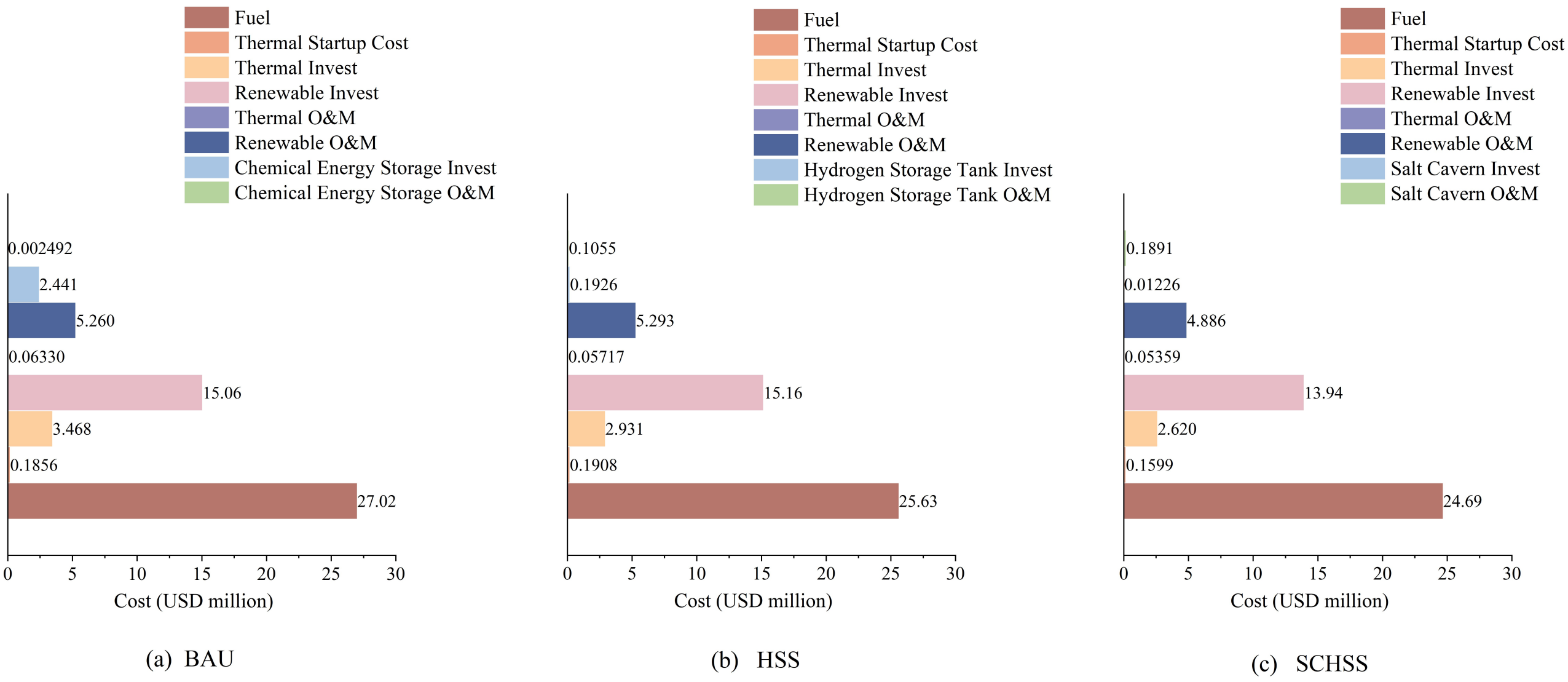

***Fig.17*** *Distribution of power system operation cost at 50% penetration*

The distribution of the power system operation costs in three different energy storage scenarios at 50% penetration is shown in Fig.17. The fuel cost, investment cost and operation costs of thermal units in the SCHSS scenario are the lowest among the three scenarios. In addition, the investment and operation costs of variable renewable energy units in the SCHSS scenario are also the lowest. From the data in Fig.17, in the SCHSS scenario, the investment and operation costs of the energy storage are 91.7% lower than that in the BAU scenario, and 32.5% lower than that in the HSS scenario. The overall cost of SCHSS-incorporated power system will be reduced by 22.2% compared with BAU scenario, and reduced by 6.3% compared with HSS scenario. This shows that salt cavern hydrogen storage has good economics, and helps to reduce the overall operation cost of the power system.

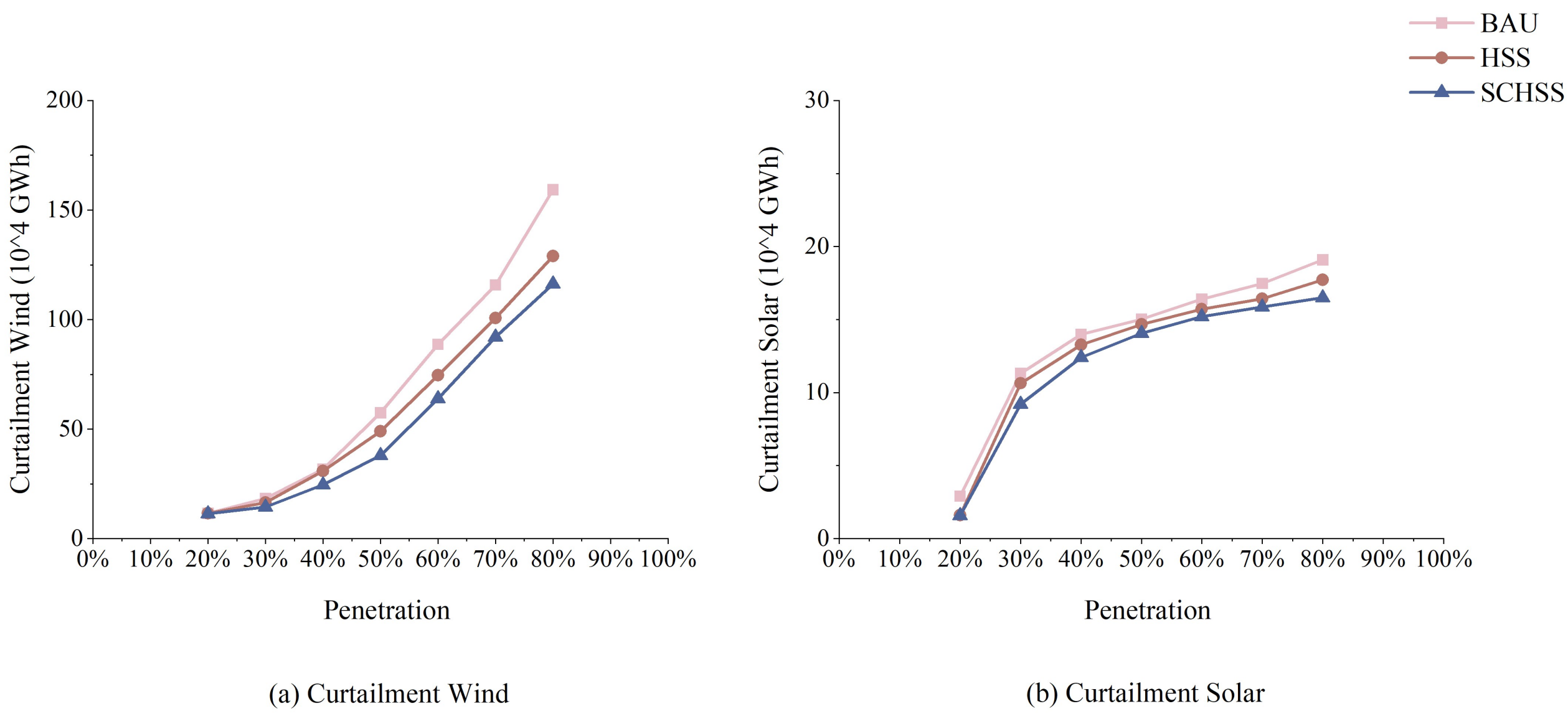

***Fig.18*** *Wind and solar curtailment at different renewable energy penetration rates*

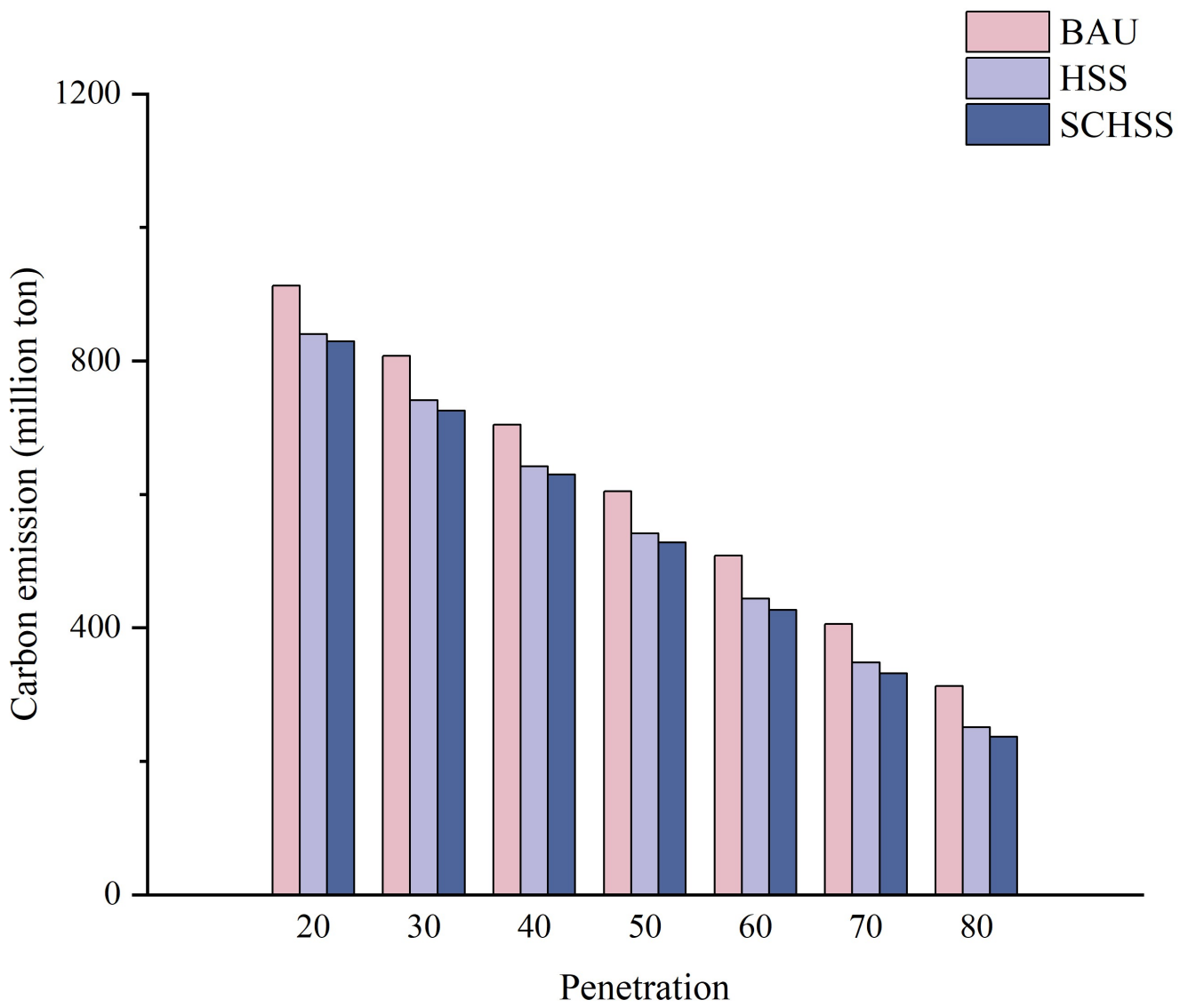

***Fig.19*** *Carbon emissions at different renewable energy penetration rates*

Fig.18 shows the curtailment of wind and solar power under different variable renewable energy penetration rates. As the penetration increases, wind and solar curtailment increases in all three scenarios. The wind curtailment and solar curtailment in the SCHSS scenario is lower than that in the HSS and BAU scenarios. When the variable renewable energy penetration rate is 80%, compared with BAU scenario, the wind curtailment in the SCHSS scenario is reduced by 27.0%, and the solar curtailment is reduced by 13.6%. The amount of the overall curtailed renewable power will be reduced by 25.6%. Compared with HSS scenario, the wind curtailment in the SCHSS scenario has been reduced by 9.8%, and the solar curtailment is reduced by 6.9%. The amount of the overall curtailed renewable power will be reduced by 9.5%.

Fig.19 indicates the carbon emissions at different variable renewable energy penetration rates. Carbon dioxide emissions decrease in all three scenarios as variable renewable energy penetration increases. Among them, the SCHSS scenario has the least carbon emissions. When the variable renewable energy penetration rate is 80%, the carbon emissions under the SCHSS scenario are 24.4% lower than that in the BAU scenario, and 5.8% lower than the HSS scenario.

### *4.3 Fast unit combined*

Fig. 20 demonstrate the solution results of the new capacity composition of different units under the fast unit combination method and the traditional modeling method. Fig. 21 depicts the solution time under the fast unit combination method and the traditional modeling method. According to the simulation results, there is no significant difference between the two solution results. For the short simulation period, the fast unit combination solution time is much shorter than the traditional integer modeling method. For the long simulation period, the traditional integer planning method cannot be solved due to the large amount of data. Therefore, only the fast unit combination method proposed in this article can meet the needs of the long - term period simulation.

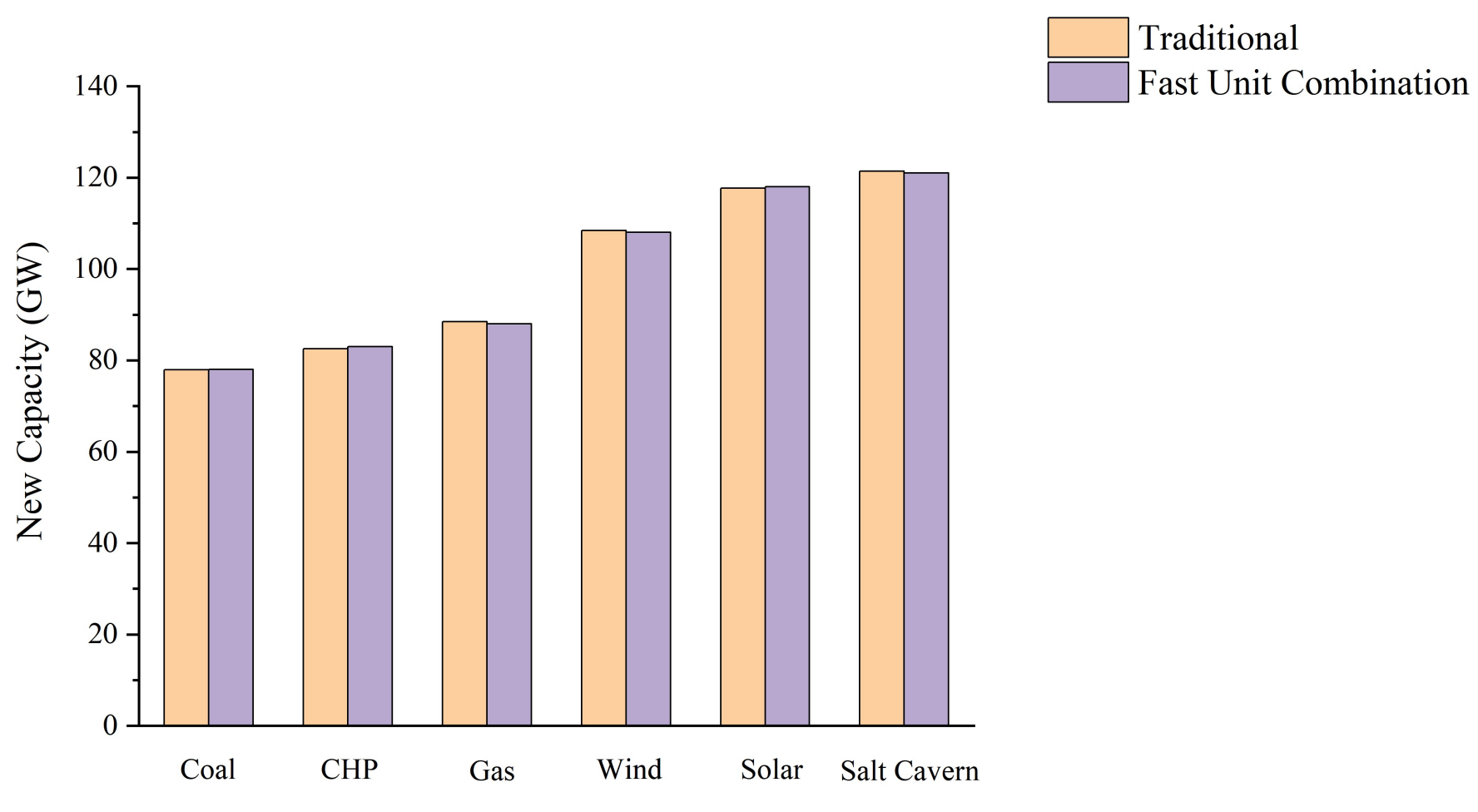

***Fig.20*** *Accuracy verification of fast unit combination*

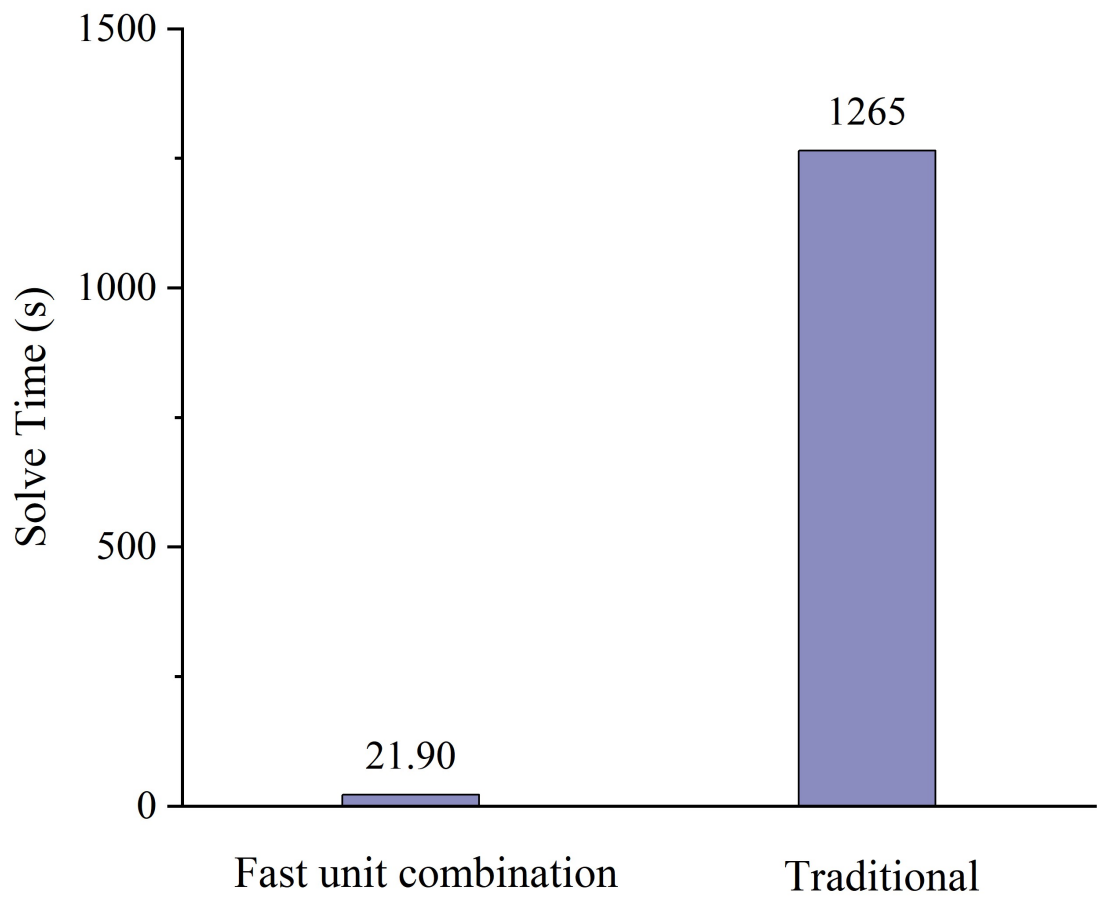

***Fig.21*** *Solution speed verification of fast unit combination*

**5. CONCLUSION**

In order to use the hydrogen energy for seasonal peak regulation, this paper establishes a long-term scale power system planning model considering the salt cavern hydrogen storage and the fast unit combination method.

The model is applied to Jiangsu Province, China for a one-year time-scale simulation, and the results show that: With a higher renewable penetration level, SCHSS takes more advantages in both economy and environmental friendliness. Compared with the traditional electrochemical energy storage method, the overall cost of the SCHSS-incorporated power system will be reduced by 22.2% with the carbon emission reduction of 24.4% (80% renewable penetration). The amount of the curtailed renewable power will be reduced by 25.6% at most (80% renewable penetration). Compared with the energy storage method using hydrogen storage tanks, the overall cost of the SCHSS-incorporated power system will be reduced by 6.3% with the carbon emission reduction of 5.8% (80% renewable penetration). The amount of the curtailed renewable power will be reduced by 9.5% at most (80% renewable penetration). Using SCHSS as an energy storage method for power systems can effectively improve the economics of the power systems with high renewable energy penetration, and improve the utilization rate of the variable renewable energy. It is of great significance to reduce the carbon emission in power systems.

In the three scenarios in this paper, the SCHSS-incorporated power system has the highest newly-built capacity of the energy storage, and the newly-built capacity of thermal power units is the lowest. Since the levelized cost of the hydrogen storage in salt caverns is the lowest, more energy storage can be introduced at low cost for peak regulation and replacing thermal power units to increase the flexibility of the power system. In addition, with the increase of the penetration of variable renewable energy, the newly-built capacity of renewable energy has risen rapidly, among which the newly-built capacity of wind power is the largest. This is because the average capacity factor of wind power is greater than that of solar power, and it is more economical to increase wind power capacity.

Furthermore, a fast unit combination method is applied to the model, which can effectively improve the solution speed. It can also overcome the disadvantage that traditional integer methods cannot solve the long-period simulation. Simulations show that the method has a high accuracy.

## 6. ACKNOWLEDGMENT

This work was supported by National Natural Science Foundation of China - Enterprise Innovation and Development Joint Fund (U22B20102).